\documentclass[11pt,a4paper]{article}
\let\oldtitle=\title
\renewcommand{\title}[1]{\oldtitle{\vskip -90pt
{\begin{normalsize}\begin{flushright}
chao-dyn/9906040 \\THU-99/17\\ June 1999\\ 
\end{flushright}\vskip 70pt
\end{normalsize}}
\bf\Large #1}} 

\usepackage{psfig}

\begin{document}

\title{Kinetic Theory of Dynamical Systems}
\author{R. van Zon and H. van Beijeren\\
\normalsize \it Institute for Theoretical Physics, University of Utrecht, \\
\normalsize \it Princetonplein 5, 3584 CC Utrecht, The Netherlands\\
{}\\
J.\ R.\ Dorfman\\
\normalsize\it
Institute for Physical
Science and Technology, and Department of Physics, \\ 
\normalsize\it
University of
Maryland, College Park, MD, 20742, USA.}

\date{June 24, 1999}

\maketitle 

\begin{abstract}

It is generally believed that the dynamics of simple fluids can be considered
to be chaotic, at least to the extent that they can be modeled as classical
systems of particles interacting with short range, repulsive forces. Here we
give a brief introduction to those parts of chaos theory that are relevant for
understanding some features of non-equilibrium processes in fluids. We
introduce the notions of  Lyapunov exponents, Kolmogorov-Sinai  entropy and
related quantities using some simple low-dimensional systems as ``toy" models
of the more complicated systems encountered in the study of fluids. We then
show how familiar methods used in the kinetic theory of gases can be employed
for explicit,  analytical calculations of  the largest Lyapunov exponent and KS
 entropy for dilute gases composed of hard spheres in $d$ dimensions. We
conclude with a brief discussion of interesting, open problems.

\end{abstract}

\newcommand{\vect}[1]{\vec{#1}}
\newcommand{\matr}[1]{\mbox{\boldmath $#1$}}
\newcommand{\eql}[1]{\label{eq:#1}}
\newcommand{\eq}[1]{Eq.~(\ref{eq:#1})}
\newcommand{\eqlabel}[1]{\eql{#1}}
\newcommand{\figl}[1]{\label{fig:#1}}
\newcommand{\dd}[2]{\frac{\partial #1}{\partial #2}}
\newcommand{\ddh}[2]{\frac{d #1}{d #2}}
\newcommand{\dif}{\delta}
\newcommand{\vvrp}{\vec{r}^{\,\prime}}
\newcommand{\vvp}{\vec{v}^{\,\prime}}
\newcommand{\vvr}{\vec{r}}
\newcommand{\vv}{\vec{v}}
\newcommand{\vdrp}{\dif\vec{r}^{\,\prime}}
\newcommand{\vdvp}{\dif\vec{v}^{\,\prime}}
\newcommand{\vdr}{\dif\vec{r}}
\newcommand{\vdv}{\dif\vec{v}}
\newcommand{\vhat}[1]{\hat{#1}}
\newcommand{\fig}[1]{Fig.~\ref{fig:#1}}
\newcommand{\identity}{\matr{1}}
\newcommand{\ho}{\mbox{h.o.}}
\newcommand{\Reg}{\gamma_R}
\newcommand{\Img}{\gamma_I}
\newcommand{\dg}{\delta\gamma}
\newcommand{\Redg}{\dg_R}
\newcommand{\Imdg}{\dg_I}
\newcommand{\dw}{\delta w}
\newcommand{\citeref}[1]{Ref.~\cite{#1}}
\newcommand{\citerefs}[1]{Refs.~\cite{#1}}

\section{Introduction}

 We consider here a classical many-particle system, a gas of hard spheres or of
hard disks. Our principal concern will be to develop methods by means of which
we can understand and calculate the properties of such gases as chaotic
dynamical systems. It is, of course, well known that to describe the
macroscopic, equilibrium properties of such gases, we can easily dispense with
any knowledge of most of the dynamical properties of the particles of which the
gas is composed. That is, one can use thermodynamics  and equilibrium
statistical mechanics, i.e. statistical thermodynamics, to describe the
relevant equilibrium properties of the gas.  All of the relevant microscopic
properties of the system needed for statistical thermodynamics are contained in
the partition sum, which is defined in terms of the  Hamiltonian of the system.
The partition function is based on a probability measure on phase space.  The
macroscopic properties are simply related to averages with respect to this
measure, of certain microscopic expressions. Of course it is far from trivial
to compute these averages for anything like a real physical system.

Our interest here, though, is to consider a gas as a mechanical system and to
understand its behavior in time, rather than its equilibrium properties, and to
try to make quantitative statements about the  motion of the trajectories of
the phase points that describe the gas, in the usual $2Nd$-dimensional
phase-space, $\Gamma$-space, of the system, where  $N$ is the number of
particles, $d$ the number of the spatial dimensions of the system, $d=2$ or
$3$, and the phase-space has $dN$ spatial coordinates and $dN$ momentum
coordinates. We take the particles to be identical hard spheres or hard disks,
each of mass $m$ and diameter  $\sigma$. When we wish to describe the typical
or average properties of the system, we must start with the  specification of
some useful probability measure, with respect to which averages can be defined.
Any dynamical system, therefore, consists of: 1) a space $\Gamma$, 2) a measure
$\mu(A)$, $A\subset\Gamma$, and 3) a transformation  $S:
\Gamma\rightarrow\Gamma$.  We will see that the dynamical viewpoint can explain
some features of macroscopic systems from their microscopic behavior. The
explanations can be followed most easily in dynamical systems of very low
dimensionality. However, even in simple low dimensional systems, dynamics may
become so complicated that is is effectively impossible to follow the dynamics
for long time, starting from a typical initial point, and we will be forced to
consider typical behaviors using some appropriate probability measure. Our
interest will be focused on chaotic systems which have the property that any
uncertainty in the specification of the exact initial state of the system will
grow exponentially in time, to the point where the future of a phase-space
point can no longer be predicted to within a reasonable accuracy\cite{Ott}. But
we can still say  something about probabilities.

It turns out that there is a close connection between the chaoticity of the
system and issues like irreversibility on the macroscopic level and, for a gas
of particles that interact with short-range forces, the validity of kinetic
theory\cite{DorfmanBeijeren}. This connection will first be outlined in section
\ref{sec:dynsys}, for low dimensional systems. A more extensive treatment can
be found in \citeref{Dorfmanbook}. In section \ref{sec:lyap} we return to a
high dimensional system in the form of a hard sphere gas in equilibrium. At low
densities we can use kinetic theory to calculate a measure of chaoticity called
the largest Lyapunov exponent. In section \ref{sec:hKS} another chaotic
characteristic of this system is calculated using kinetic theory: the
Kolmogorov-Sinai entropy. In section \ref{sec:conclusion} we make some
concluding remarks and present some open questions.

\section{Dynamical Systems}
\label{sec:dynsys}

The standard approaches to the theory of  non-equilibrium processes in fluids
are based on three foundational pillars: (1) The identification of the
macroscopic quantities of physical interest as averages of microscopic
quantities over an appropriate ensemble of similarly prepared systems; (2) The
use of the Liouville equation, either in its classical or  in its quantum
mechanical version, to compute the time evolution of the ensemble distribution
function; and (3) The utilization of some kind of physically reasonable
factorization assumption for the ensemble distribution function in order to
transform the Liouville equation into a tractable equation whose solution can
be used to make quantitative statements about the macroscopic quantities. Such
a procedure is followed in the derivation of the Navier-Stokes fluid dynamic
equations from the Liouville  equation\cite{erndorf2} for general fluids, and
in the derivation of the Boltzmann transport equation, and its extensions to
higher densities,  from the Liouville equation for dilute and moderately dense
gases. More  phenomenological approaches to irreversible behavior in fluids
often depend on explicit stochastic  assumptions about the underlying dynamical
processes taking place  in the fluid\cite{wax}.

While they are of the highest importance for the development of theories of
irreversible processes in fluids, both of these approaches to irreversible
behavior leave the answers to some fundamental questions obscure.  In
particular, these approaches offer only qualitative insights into the reasons
for the validity of the stochastic assumptions imbedded in these various
procedures -- either through factorization assumptions, which in essence, are
statements about correlations and probabilities -- or through the replacement
of the exact dynamics by a stochastic, Langevin-type, dynamics. Further, while
the approaches outlined above do predict an approach to an equilibrium state,
under the proper physical conditions, and do provide experimentally verifiable
statements about the approach to equilibrium, they do not give a complete
picture of why the system approaches an equilibrium state, based upon the
underlying microscopic dynamics. The general arguments for the use of
stochastic methods are based upon the randomness of the microscopic motions of
the particles but are not much more specific than that. The picture that we
have  of the approach to  equilibrium is generally based on the idea that the
local averages of conserved mechanical quantities, such as mass, momentum, and
energy,  change very slowly in time compared to the local averages of
nonconserved quantities. Thus the  macroscopic behavior will be dominated by
the slowest variables in the system, the local conserved quantities, and the
equilibrium state will be achieved when these quantities have reached steady,
homogeneous values. This picture, suggested by solutions of the Boltzmann
equation, has led to important advances in the theory of fluids, among others,
to mode-coupling theory. What is missing from it is a basic understanding of
the necessary (or sufficient) properties of the intermolecular potential for
the system to approach an equilibrium state, as well as an understanding of the
properties of the trajectories of the system, and the evolution of measures in
$\Gamma$-space,  that are responsible for the approach to equilibrium states
and, perhaps, under different boundary conditions, to more complicated, but
interesting non-equilibrium steady states.

The application of ideas from dynamical systems theory to non-equilibrium
statistical mechanics allows us to make some progress in resolving the issues
described above. The application of ideas from chaos theory, in particular,
enables us to make some quantitative statements about the type and degree of
randomness of a dynamical system, even of large systems typically treated by
statistical mechanics. It also allows us to describe equilibrium and
non-equilibrium states  of a system in terms of probability measures defined in
$\Gamma$-space, and in terms of the time evolution of these measures. Moreover,
there are interesting and unexpected connections between the macroscopic
transport coefficients that describe the approach of a fluid to equilibrium,
and microscopic  quantities that describe the chaotic behavior of a fluid,
considered as a large, dynamical system. In this section we will outline some
of these rather new ideas, and illustrate their applications to statistical
mechanics by seeing how they work for systems of low dimensions and then
generalizing them, when possible, to higher dimensional systems. We begin with
a very simple two-dimensional reversible system, the baker's map, which
exhibits many of the  features we would like to see in more general, higher
dimensional systems.

\subsection{The baker's map}

\begin{figure}[t]
   \centerline{\psfig{figure=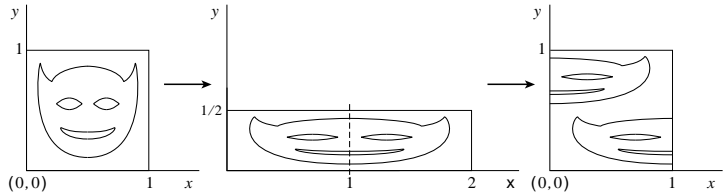,width=9.5cm}}
   \caption{
            The baker's transformation.
           }
   \label{fig:baker}
\end{figure}

The simplest example of a reversible system with chaotic dynamics is probably
the baker's map. Here we consider a two-dimensional phase space on a unit
square. That is, $\Gamma = (x,y); 0\leq x,y \leq 1$. The map, ${\bf B}$,
operates only at discrete time steps, and moves points $(x,y)$ to $ {\bf
B}(x,y)=(x',y')$ given by

\begin{eqnarray}
{\bf B}\left(\begin{array}{c}
x \\
y
\end{array}\right) =
\left( \begin{array}{c}
 x' \\
 y'
\end{array} \right) & = &\left( \begin{array}{c}
                                2x  \\
                               y/2   \end{array}\right) \,\,\,\,\,
{\mbox{for}}\,\,
0 \leq x <1/2; \,\,\,  {\mbox{and}} \nonumber \\
 & = & \left( \begin{array}{c}
    2x-1  \\
     (y+1)/2  \end{array}\right) \,\,\,\,\, {\mbox{for}}\,\, 1/2 \leq x <1.
\eqlabel{bm1}
\end{eqnarray}
This map is illustrated in  \fig{baker}.  It is immediately clear  that this
map possesses an inverse, ${\bf B}^{-1}$, given by
\begin{eqnarray}
{\bf B}^{-1}\left(\begin{array}{c}
x \\
y
\end{array}\right) & = &\left(\begin{array}{c}
x/2 \\
2y
\end{array}\right)\,\,\,\,\,{\rm for}\,\,\, 0 \leq y < 1/2\,\,\,{\mbox{and}}
\nonumber \\
& =& \left(\begin{array}{c}
(x+1)/2 \\
2y-1
\end{array}\right)\,\,\,\,\,{\mbox{for}}\,\,\, 1/2 \leq y <1.
\eqlabel{bm2}
\end{eqnarray}

The baker's map is clearly area-preserving, and it is time-reversible in the
sense that the  transformation ${\bf T}:(x,y)\rightarrow(1-y,1-x)$ serves as a
time reversal transformation for this map, such that ${\bf T}\circ{\bf
B}\circ{\bf T} = {\bf B}^{-1}$, and ${\bf T}\circ{\bf T}={\bf 1}$, where ${\bf
1}$ is the unit operator.

Now we regard the unit square as a ``toy" phase-space. The dynamics of the
baker's map in this phase-space has the following properties: 1) Consider two
infinitesimally separated points. Unless they have exactly the same
$x$-coordinates, or the same $y$-coordinates, the images of these two points
under several applications, called iterations, of the map ${\bf B}$ will cause
the  $x$-components to separate {\em exponentially} with the number of
applications, with an exponent of $ \lambda_{+}=\ln2$, whereas the
$y$-components will converge exponentially to a common value with an exponent
of $\lambda_{-}=-\ln2$. The exponents, $\lambda_{\pm}$, characterizing
exponential  separation or convergence of points in phase-space are called {\em
Lyapunov exponents}. The  directions in which the points converge
exponentially,  in this case just the $y$-direction, are called {\em stable}
directions, and the directions in which they separate exponentially, in this
case the $x$-direction, are called {\em unstable} directions. The fact that the
Lyapunov exponents sum to zero is a simple consequence of the area preserving
property of the baker's map, as can easily be seen by considering the
evolution, with successive iterations, of the small rectangle with corners at
$(x,y),(x+\delta x,y),(x,y+\delta y),(x+\delta x,y+\delta y)$ under the baker
map, ${\bf B}$. This rectangle has constant area, but grows exponentially long
in the $x$-direction, and exponentially thin in the $y$-direction. Sooner or
later it gets stretched and folded in such a way that on a coarse grained
scale, the unit square is covered uniformly. We will describe this behavior on
a coarse-grained scale, by saying that the distribution of points becomes
weakly uniform. It is important  to note that the projection of the small
rectangle onto the $x$-axis will be uniform in a time, $n_{u}$ on the order of
\begin{equation}
 n_{u} \sim \frac{-\ln \delta x}{\lambda_{+}},
\eqlabel{bm3}
\end{equation}
where $\lambda_{+} = \ln 2$, is the positive Lyapunov exponent for the baker's
map. That is, the projection of the small rectangle on the unstable direction
becomes uniform much sooner than the distribution of points on the entire unit
square becomes weakly uniform.

\subsection{The Arnold cat map and hyperbolic systems}

\begin{figure}[t]
   \centerline{\psfig{figure=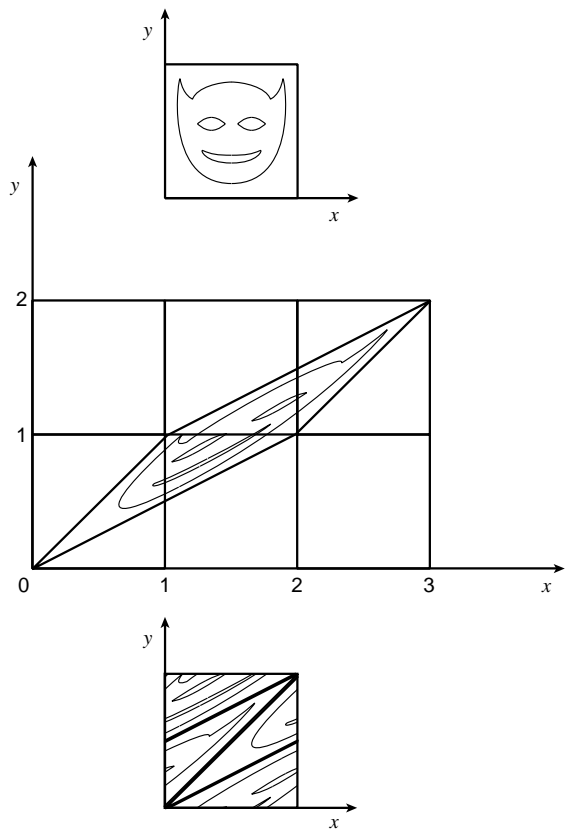,height=11cm}}
   \caption{
            The Arnold cat map.
           }
   \label{fig:catmap}
\end{figure}

A different map with a similar dynamical behavior, i.e., area preserving, with
exponentially separating and converging trajectories characterized by positive
and negative Lyapunov exponents, is provided by the Arnold cat map, ${\bf
T}(x,y)$, illustrated in Fig.  \ref{fig:catmap},  and given by
\begin{equation}\eqlabel{bm4}
\left(\begin{array}{c}
x' \\
y' \end{array}\right) = {\bf{T}}\left(\begin{array}{c}
x \\
y \end{array}\right) =\left(\begin{array}{cc}
2 & 1 \\
1 & 1 \end{array}\right)\left(\begin{array}{c}
x \\
y \end{array}\right) \,\,\, {\mbox{mod}} \,\,1.
\end{equation}
The area preserving property is guaranteed by the fact that the matrix
representation of ${\bf T}$ has unit determinant, the integer coefficients of
the matrix together with the mod $1$ condition implies that the unit square, or
more properly, the unit torus, is mapped smoothly onto itself by ${\bf T}$.
Such area preserving maps with integer coefficients are called {\em toral
automorphisms}. The Arnold cat map also has stable and unstable directions
associated with positive and negative Lyapunov exponents given by
$\lambda_{\pm} =\ln[(3 \pm \surd{5})/2]$. In a way much like the baker's map, a
small region of the unit square  will be stretched and squeezed under the
iterated action of the cat map, with projections on both the $x$ and $y$-axes
becoming uniform on a similar time scale as in \eq{bm3}, and with the
distribution of points becoming weakly uniform on a longer time scale.

The baker's map and the Arnold cat map are two simple examples of what are
called {\em  hyperbolic} dynamical systems. Briefly, and somewhat loosely
stated, hyperbolic dynamical systems are defined by the action of some
dynamical transformation, ${\bf S}$ on a phase-space $\Gamma$, such that: (a)
one can identify stable and unstable directions in $\Gamma$, under the action
of ${\bf S}$, with negative and positive Lyapunov exponents, all bounded away
from zero; (b) the stable and unstable manifolds (lines, surfaces, etc.) are
continuous functions of the variables that define the phase space, and when the
two manifolds intersect, they do so transversely; (c) the system is {\it
transitive}, i.e., there exists some trajectory in the phase-space that is
dense on the phase-space; and (d) for maps, i.e., for dynamical systems where
${\bf S}$ acts only at discrete times, there are no directions in phase-space
with a Lyapunov exponent of zero,  while for ``flows", i.e. systems where ${\bf
S}$ depends upon a continuous time parameter, the only direction in $\Gamma$
with a zero Lyapunov exponent is the direction of the flow. Clearly the baker's
map and the Arnold cat map are  hyperbolic maps. A typical flow that one might
examine for hyperbolicity is the motion of a phase point on surfaces of
constant energy for a system of interacting particles.  \subsection{Ergodic and
mixing systems}

The baker's map and the Arnold cat map are also examples of dynamical systems
which are {\it ergodic} and {\it mixing}. Ergodicity is the property of a
dynamical system that the time average of any integrable function of the
phase-space variables will be equal to the ensemble average of this function,
the average taken with respect to an appropriate  time translation invariant,
equilibrium measure. That is, if $f(\Gamma)$ is an integrable function, then
\begin{equation}
\lim_{n\to\infty}\frac{1}{n}\sum_{j=o}^{n-1}f({\bf S}^{j}\Gamma) =\int
f(\Gamma)\,\mu(d\Gamma),
\eqlabel{bm5}
\end{equation}
where $\mu(A)$ is a measure that is {\it invariant}, i.e., $\mu(A) =\mu({\bf
S}^{-1}A)$ for any non-trivial set $A$, and {\it ergodic}, meaning that it is
impossible to divide the whole phase-space into two invariant sets, each of
positive measure\footnote{\eq{bm5} doesn't have to hold for all points
$\Gamma$, as long as the set of points violating it has measure zero, with
respect to  the measure in the definition of the dynamical system.}. It is
generally assumed, but not always proved, that our systems possess a unique
ergodic measure. Students of statistical mechanics will naturally associate the
idea of an ergodic system with the name of Boltzmann who used this idea to base
equilibrium statistical mechanics on the laws of
mechanics\footnote{Traditionally, statistical mechanical systems were called
ergodic if they are ergodic under Hamiltonian flow with the Liouville measure
on the energy shell. }.

Much of equilibrium statistical mechanics can be based on the laws of large
numbers, and strict ergodicity, in the sense of Boltzmann, is not that
essential. However,  non-equilibrium statistical mechanics requires some deep
underpinnings from mechanics, or from the theory of stochastic processes. Here
we take the point of view that Hamiltonian mechanics is all that is needed,
but that is certainly not the only possible point of view. For  non-equilibrium
statistical mechanics, it is useful to explore an idea of Gibbs, which is
called the mixing property of a dynamical system. Mixing systems are always
ergodic, but the reverse is not always true.  To define a mixing system, we
consider two arbitrary sets in the phase-space, $A$ and $B$, say, both of
nonzero measure, and  the evolution of the set $A$ in time. Suppose after $n$
iterations of the map ${\bf S}$ the set $A$ has moved to ${\bf S}^{n}A$, then
the system is mixing if
\begin{equation}
\lim_{n\to\infty}\frac{\mu(B\cap{\bf S}^{n}A)}{\mu(B)}
=\frac{\mu(A)}{\mu(\Gamma)},
\eqlabel{bm6}
\end{equation}
where $\mu(\Gamma)$ is the measure of the entire phase-space, such as the unit
square for the baker's map or the cat map, or the constant energy surface for a
more general system. The mixing condition simply means that the time evolution
of a set in phase-space is such that, in a coarse grained sense, it gets
uniformly distributed, with respect to the measure $\mu$, over the entire
phase-space. It can be proved rather easily that for a mixing system,
non-equilibrium averages of integrable functions $f$ will approach their
equilibrium values in the course of time.

\subsection{The approach to equilibrium}

A nice illustration of the approach to equilibrium, as provided by baker or cat
maps, is to consider the behavior in time of reduced distribution functions.
That is, if we think of the unit square, again, as a phase space, then we can
define a phase-space distribution function, $\rho_{n}(x,y)$, as a function of
the number of iterations of the map, $n$, and the coordinates $x$ and $y$. The
phase-space distribution function satisfies a discrete-time version of the
Liouville equation, which is a form of the Frobenius-Perron equation for area
preserving maps. The appropriate equation is
\begin{equation}
\rho_{n}(x,y) = \rho_{n-1}({\bf B}^{-1}(x,y)),
\eqlabel{bm7}
\end{equation}
which, written out in full detail, becomes
\begin{eqnarray}
\rho_{n}(x,y)& =& \rho_{n-1}(x/2, 2y) \,\,\,{\mbox{for}}\,\,0\leq y <1/2
\nonumber \\
    &=& \rho_{n-1}((x+1)/2, 2y-1) \,\,\,\, {\mbox{for}}\,\, 1/2 \leq y
<1.
\eqlabel{bm8}
\end{eqnarray}
A similar, but somewhat more complicated equation could be given for the Arnold
cat map, but we will not use it here.

For these simple two dimensional models, a reduced distribution function is
obtained by integrating the distribution function over one of the two
phase-space variables, $x$ or $y$. This integration is motivated by the fact
that for a system of $N$ particles, we are not particularly interested in the
full $N$-particle distribution function, but rather in the one or two-particle
distribution functions that can be used to evaluate the macroscopic quantities
of interest, such as mass, momentum, and energy densities. Since our simple
maps have only two coordinates, we can only consider the very simple case where
a reduced distribution function is obtained by integrating over one of the
coordinates. For the baker's map we will  construct the distribution function
for the density of points in the $x$-direction, for reasons that will become
clear as we proceed. That is, we define a reduced distribution function
$W_{n}(x)$ by
\begin{equation}
W_{n}(x)=\int_{0}^{1}dy\,\rho_{n}(x,y).
\eqlabel{bm9}
\end{equation}
Using \eq{bm8}, we can easily obtain a difference equation for $W_{n}(x)$, as
\begin{equation}
W_{n}(x) = \frac{1}{2}\left[W_{n-1}(\frac{x}{2}) +
W_{n-1}(\frac{x+1}{2})\right].
\eqlabel{bm10}
\end{equation}

\begin{figure}[t]
        \centerline{\psfig{figure=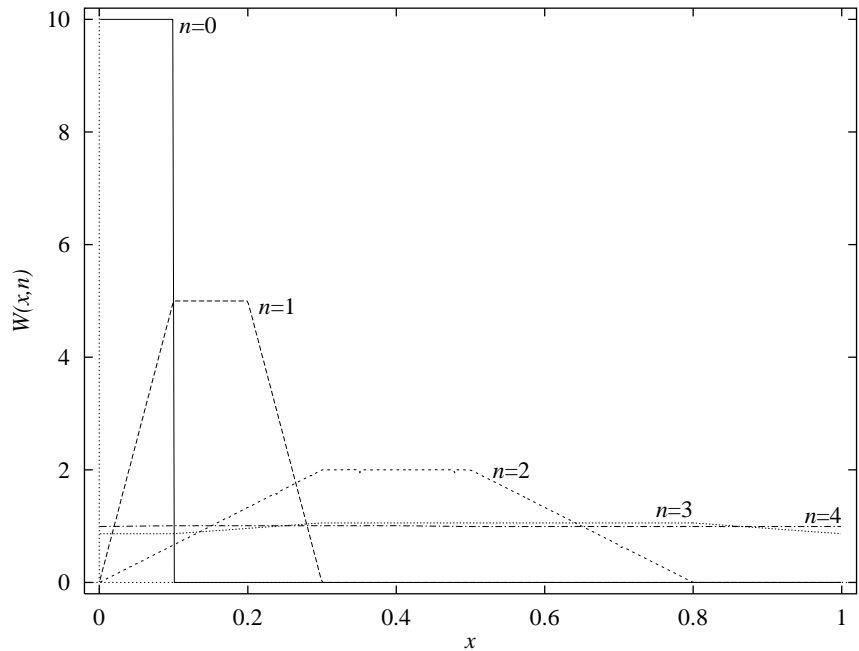,width=\textwidth}}
   \caption{
            The projection, $W(x,n)$, onto the $x$-coordinate,
            of the phase-space distribution function for the
            Arnold cat map.
           }
   \label{fig:arnold2}
\end{figure}

\begin{figure}[th]
   \centerline{\psfig{figure=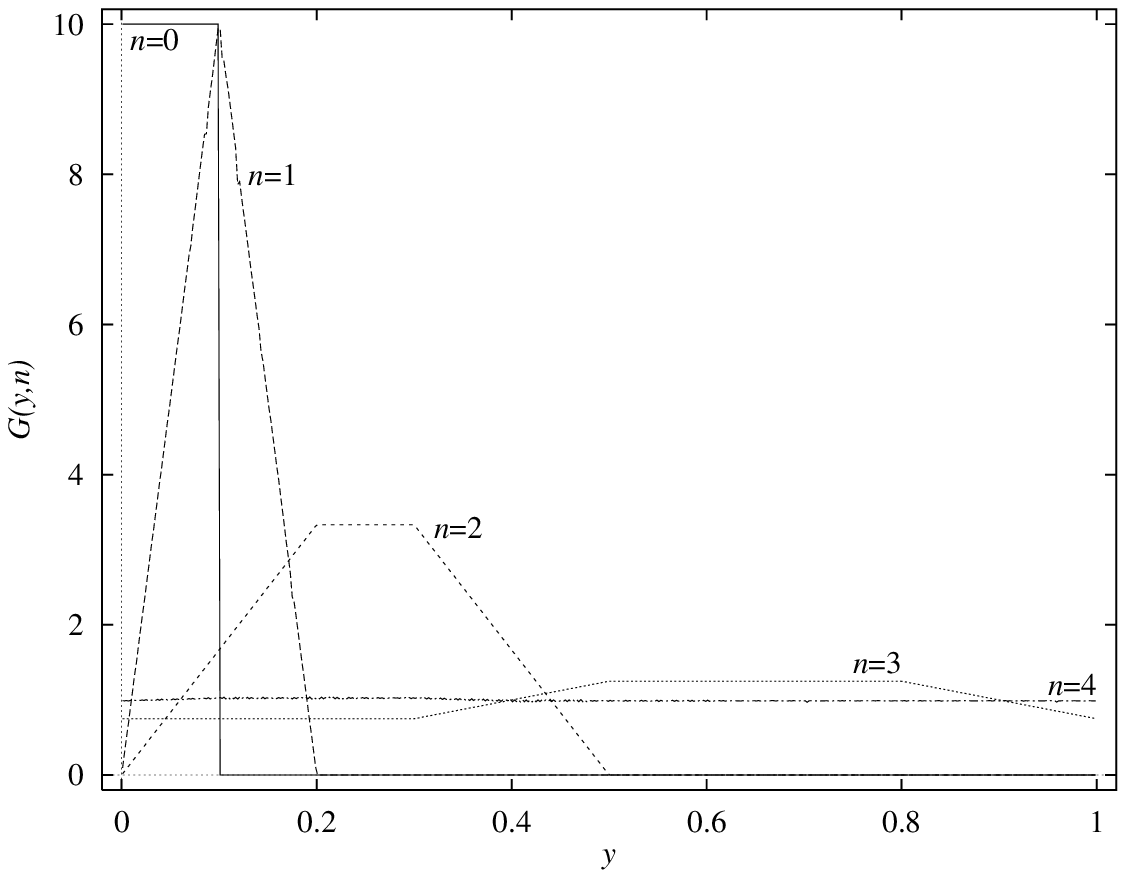,width=\textwidth}}
   \caption{
            The projection, $G(y,n)$, onto the $y$-coordinate,
            of the phase-space distribution function for the
            Arnold cat map.
           }
   \label{fig:arnold3}
\end{figure}

This equation is, among other things, the Frobenius-Perron equation for the
one-dimensional map, $x'= 2x$ (mod $1$), on the interval $(0,1)$. What is more
important here, though, is that except for very special initial values for
$W_{0}(x)$, such as Dirac delta functions on the periodic points of the map,
$W_{n}(x)$ approaches a constant, independent of $x$, as $n\to\infty$. This may
be proved in a number of ways, but may be understood most simply by just
drawing some possible functional forms for $W_{0}(x)$ and follow what happens
to them after a few iterations of \eq{bm10}. A standard procedure is to make a
Fourier expansion of $W_{0}(x)$, and to notice that only the constant term
remains as the number of iterations gets large. The approach to equilibrium in
this simple system can be associated with the properties of the expanding
manifold in our simple two-dimensional phase-space. Because of the stretching
of regions in phase-space in the unstable directions, functions defined on the
unstable manifold will get ``smoothed out" in the course of time, much the same
way that a ball of dough gets smoother and smoother along the direction that
the baker stretches it. The initial wrinkles in the phase-space distribution
function will not get smoothed out along the stable direction, on the contrary,
they typically will get more and more wrinkled as the system evolves. From
these considerations we can see that the integration of the phase-space
distribution over the stable direction in \eq{bm9} was not chosen accidentally;
 had we integrated over $x$ instead, we would not have obtained an equation
with a nice equilibrium solution as $n\to\infty$. In fact, a typical initial
distribution will become smooth in the expanding directions but very striated
in the contracting  directions. However, eventually it will look uniform on a
coarse grained scale, consistent with the mixing behavior of the baker's map.

The connection between the approach to equilibrium and the expanding direction
of a measure preserving map can be further explored by considering the Arnold
cat map. Here, the expanding direction is along a line that is not aligned
along either of the coordinate axes. One would expect that for this model a
projection of the phase-space distribution function along either the $x$-axis,
or the $y$-axis, would approach an equilibrium value. That this is so can be
seen from a simple computer calculation. We start with a phase-space
distribution that is concentrated in a small region $0 \leq x,y \leq 0.1$. We
then follow the evolution in time of $x$ and $y$ projections of the
distribution function. In  Figs. \ref{fig:arnold2} and \ref{fig:arnold3} we can
easily see that these distribution approach constant values after  three or
four iterations of the map, much before the entire phase-space distribution is
smooth, even on a reasonably coarse grained scale, which for this arrangement
takes  eight to  ten iterations. This observation may be generalized: reduced
distribution functions on some lower dimensional projection of phase-space will
always become smooth under the dynamics, unless the projected space is entirely
spanned by stable directions (hence is some subset of the stable manifold).

\begin{figure}[ht]
   \centerline{\psfig{figure=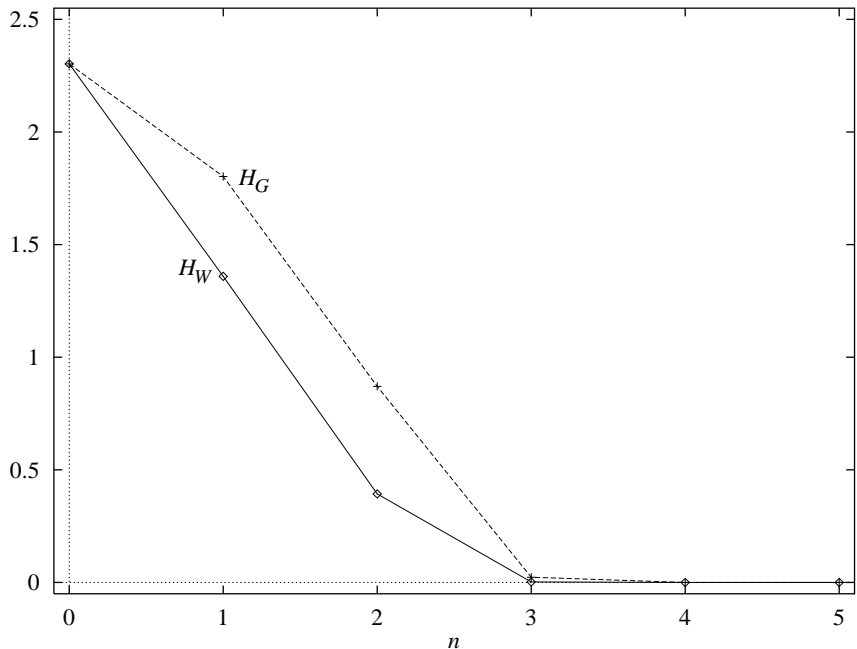,width=\textwidth}}
   \caption{
            The Boltzmann $H$-functions $H_W(n)$ and $H_G(n)$,
            \index{Arnold cat map!$H$-function}
            obtained by using $W(x,n)$ and $G(y,n)$ respectively.
           }
   \label{fig:arnoldH}
\end{figure}

Not only does one see an approach to an equilibrium distribution for the
projected distribution functions for these maps, one also sees that a suitably
defined  Boltzmann $H$-function decreases monotonically as the number of
iterations increases. This is illustrated in  Fig. \ref{fig:arnoldH} for the
Arnold cat map, for both projected distribution functions, starting from the
same initial state as described above. The figure shows the $H$-function for
both projections, as calculated on a computer. For the baker's map, we can
easily show  the monotonic decrease in the $H$ function analytically. To do
this we define the $H$-function by
\begin{equation}
H(n) =\int_{0}^{1}dx\, W_{n}(x)\,\ln W_{n}(x).
\eqlabel{bm11}
\end{equation}
If we now use the recursion relation for $W_{n}(x)$, \eq{bm10}, we find that
\begin{eqnarray}
&&\!\!\!\!H(n+1) =\int_{0}^{1}dx\,W_{n+1}(x)\,\ln W_{n+1}(x) \nonumber \\
&&\!\!\!\!\quad =
\int_{0}^{1}dx\frac{1}{2}\left[W_{n}(\frac{x}{2})+W_{n}(\frac{x+1}{2})\right]
\ln\left\{
\frac{1}{2}\left[[W_{n}(\frac{x}{2})+W_{n}(\frac{x+1}{2})\right]\right\}
\nonumber
\\
&&\!\!\!\!\quad \leq \frac{1}{2}\int_{0}^{1}dx\,\left[W_{n}(\frac{x}{2})\ln
W_{n}(\frac{x}{2})+W_{n}(\frac{x+1}{2})\ln W_{n}(\frac{x+1}{2})\right]
\nonumber \\&&\!\!\!\!
\quad =  H(n).
\eqlabel{bm12}
\end{eqnarray}
The inequality in \eq{bm12} follows from the fact that $f[(a+b)/2] \leq
[f(a)+f(b)]/2$ if $f(x)=x\ln x$. That is, a chord connecting two points on the
curve $x \ln x$ lies above the curve. Thus we see that the $H$ function
decreases with time until $W_{n}(x)$ becomes a constant.

We conclude this discussion with a few remarks. For the baker's map and the
Arnold cat map, admittedly ``toy" models, highly simplified versions of $N$
particle systems, with almost trivial phase-spaces, we have been able to derive
irreversible equations and $H$-theorems with a minimum of assumptions. We have
associated the approach to equilibrium of projected distribution functions with
the existence of unstable manifolds for the dynamics in phase-space, and the
fact that the projection is not orthogonal to the unstable directions. In a
more general context, such as a corresponding, but not yet possible, dynamical
derivation of the Boltzmann transport equation, we would expect that the
approach to equilibrium, seen here for baker and Arnold cat maps, would
correspond  to the approach to a local equilibrium state in the fluid. In such
a local equilibrium state, the system has equilibrium values for density,
temperature, and local mean velocity over distances of a few mean free paths.
Then much slower hydrodynamic processes with a different kind of dynamics
govern the approach to an overall equilibrium state for the entire
fluid.\footnote{We wish to insert one word of caution about this picture. It
seems clear from an examination of diffusion in some non-chaotic systems, such
as the famous   wind-tree model, that chaoticity is sufficient, but not always
necessary for understanding the approach to equilibrium of systems of many
particles. However, in such non chaotic systems there often is a non-dynamical
source of randomness, as in the random locations of scatterers in the wind tree
model. This non-dynamical source of randomness is not needed to explain the
approach to equilibrium in chaotic systems.}

\subsection{The Kolmogorov-Sinai entropy}

We next turn to a brief discussion of an important quantity that characterizes
both deterministic chaos of the type we have been studying, as well as Markov,
stochastic processes. This quantity is called the Kolmogorov-Sinai (KS)
entropy. For a deterministic system it measures the rate at which information
is gained about the initial state of a system. That is, suppose that we know
that the initial phase point of our system is in some small region of
$\Gamma$-space of dimension $\varepsilon$ on a side, and that we cannot resolve
the location in $\Gamma$-space to any better precision. Consider now the
evolution of this small volume element in $\Gamma$-space. For a system with
non-zero Lyapunov  exponents, this small volume will get exponentially
stretched along the expanding directions. After some time this stretching will
make some sides exponentially longer than the initial value $\varepsilon$,
typically of length $\varepsilon\exp(\lambda_{i}t)$, where $\lambda_{i}$ is one
of the positive Lyapunov exponents. Since we can resolve points in
$\Gamma$-space to within a distance of $\varepsilon$ in any direction, we can
now determine which of many small regions, of dimension $\varepsilon$ on a
side, our system is in at time $t$. Then by inferring where this region came
from in the  initial volume,  we learn more about the  initial location of the
phase point. Although it requires some careful analysis to prove, it is not
difficult to imagine that there is some direct relation between the positive
Lyapunov exponents and the KS entropy, $h_{KS}$. In fact for a closed,
hyperbolic system, Pesin has proved that the relation is as direct as our
discussion above would imply, namely,
\begin{equation}
h_{KS} = \sum_{\lambda_{i}>0}\lambda_{i}.
\eqlabel{bm13}
\end{equation}

A deep and interesting fact is that at least one way to prove Pesin's result
(which we will not do here) depends on the fact that hyperbolic systems can be
mapped onto Markov stochastic processes. That is, for such systems, with baker
and Arnold cat maps as simple examples, one can represent the dynamics to
within any arbitrary degree of precision, as a Markov process. Such Markov
processes have a measure of their own information entropy, a quantity which
measures the degree of uncertainty in the next outcome of the stochastic
process. One can show that the information entropy, suitably defined, of the
stochastic  representation of a hyperbolic dynamical system is equal to the KS
entropy of the system. This is one of the deep results of dynamical systems
theory, which provides a firm  mathematical basis  for the correspondence of
hyperbolic dynamical systems with Markov processes. It expresses the fact that
from the point of view of mathematical analysis, at least, there is no real
difference between a Newtonian, hyperbolic dynamical system with a finite KS
entropy, and a Markov stochastic process with equal values of the information
entropy. From the point of view of physics there is a big difference, of
course. The central idea of the physical approach we take is to show that
Newtonian dynamical systems are sufficiently hyperbolic  to behave as if they
were Markov stochastic systems and  in consequence all  important properties of
Markov systems apply to dynamical systems, too.

An important first step is to show that some useful models of physical systems
have positive Lyapunov exponents and a finite, positive KS entropy per
particle. In the next sections we will show how methods of statistical
mechanics can be used to calculate Lyapunov exponents and KS entropies of some
simple many-particle systems---gases of hard spheres in $d$ dimensions, where
$d = 2,3$. Before doing so, however, we briefly turn our attention to an
application of the ideas of this section to non-equilibrium statistical
mechanics.

\subsection{The escape-rate formalism for transport coefficients}

We conclude this section with a discussion of a formal relation between the
transport coefficients that characterize hydrodynamic processes in fluid
systems, and the chaotic properties, such as Lyapunov exponents and KS
entropies that characterize the underlying dynamical behavior of the fluid. The
relation of interest here is called the ``escape-rate" formula for transport
coefficients and is due to Gaspard, Nicolis, and co-workers\cite{gasnic,dogas}.
It applies to those fluids which can be considered to be classical, transitive,
hyperbolic dynamical systems. While we do not know with certainty if any models
of fluid  systems satisfy this hyperbolicity requirement, we can suppose, as a
working hypothesis, that generic fluid systems are well described as transitive
hyperbolic systems, and then explore the consequences that result. This
hypothesis has been assumed by almost all workers in this field, but  its most
satisfactory articulation was provided by Cohen and Gallavotti in their study
of  non-equilibrium fluctuations in thermostatted
systems\cite{gallco}\footnote{It is customary to use the term {\it Anosov}
system to describe a transitive hyperbolic system without singularities, such
as the Arnold cat map. The class  of Anosov systems does not include baker maps
or hard sphere systems, since discontinuities of the map or flow, present in
these systems, are not allowed by  the Anosov condition. For this reason we
prefer to generalize the chaotic hypothesis to include transitive, hyperbolic
systems.}.

\begin{figure}[t]
   \centerline{\psfig{figure=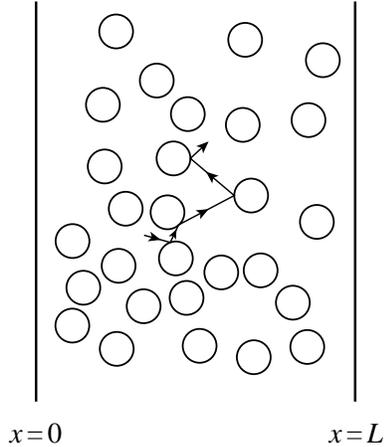,height=6.0cm}}   \caption{
            A slab geometry for diffusion in a
            system of moving particles is an array of fixed scatterers, with
absorbing
            boundaries.
           }   \label{erfig}\end{figure}

To illustrate the escape-rate formula, we consider only the case of particle
diffusion in an array of fixed scatterers, and refer to the literature for more
general cases\cite{dogas,gasbook}. We suppose that a collection of particles is
moving in a region $R$ in space, and that there is also a collection of fixed
scatterers placed in $R$, as illustrated in Fig.~(\ref{erfig}). We suppose that
the size of $R$ is  characterized  by a length $L$ which is much larger than
the typical mean free path of the particles moving in $R$. For simplicity we
suppose that the moving particles do not interact with each other but only with
the scatterers, and that the scatterers do not ``trap" the moving particles in
regions of microscopic or macroscopic size. In order to provide a
non-equilibrium situation which would exhibit the diffusive properties of this
arrangement, we suppose that $R$ is surrounded by absorbing boundaries such
that if a particle crosses the boundary of $R$ from the interior, it is
absorbed and lost to the system.

Under most circumstances \footnote{I.e., the surface area to volume of $R$
scales to zero as $L\rightarrow\infty$.}, the classical, macroscopic   dynamics
of this process is described by the diffusion equation
\begin{equation}
\frac{\partial n(\vvr,t)}{\partial t}=D\nabla^{2}n(\vvr,t),
\eqlabel{er1}
\end{equation}
where $n(\vvr,t)$ is the density of moving particles at time $t$ at point
$\vvr$, and $D$ is the coefficient of diffusion of the moving particles for
this system. The absorbing boundary conditions require that $n(\vvr,t)=0$ on
the boundaries. While the exact solution of \eq{er1} depends on the geometry of
the system, one can easily see that for long times, the total number of
particles in the system decays with time as
\begin{equation}
N(t) =\int_{R} d\vvr n(\vvr,t) \simeq N(t=0)\exp[-DA\,t/L^{2}],
\eqlabel{er2}
\end{equation}
where $A$ is a factor of order unity that depends on the geometry of the region
$R$, $D$ is the diffusion coefficient, and $L$ is the characteristic size of
$R$. We see that this macroscopic process is  characterized by an exponential
escape of particles from $R$, with  an escape-rate, $\gamma_{mac}=DA/L^{2}$.
There is a corresponding {\it microscopic} description of the escape process
based upon the properties of the trajectories of the particles moving in the
array of scatterers. There is a set of initial points in the phase space for
the moving particles which are associated with trajectories that never leave
the system for either the forward or time reversed motion. This set of points
is called a {\it repeller} and denoted by ${\cal R}$, and it is  invariant, in
the sense that any time-translation of this set of points is the set ${\cal R}$
itself. Further, the repeller usually forms a fractal set in phase space with
zero Lebesgue measure. Simple examples of repellers can be found in most books
on chaos theory\cite{Ott,Dorfmanbook}. It is possible to consider the dynamical
properties of the trajectories on the repeller, and to define the Lyapunov
exponents, $\lambda_{i}({\cal R})$, and the KS entropy, $h_{KS}({\cal R})$, of
such  trajectories\cite{gasbook,chermak,rueck}.  For transitive hyperbolic
systems, one finds that the  sum of the positive Lyapunov exponents on the
repeller is not equal to the KS entropy as would be the case for closed systems
according to Pesin's theorem,  but that the two quantities differ by an amount
equal to the rate at which the other trajectories escape from the system, which
we denote by $\gamma_{mic}$. That is
\begin{equation}
\gamma_{mic} = \sum_{\lambda_{i}>0}\lambda_{i}({\cal R}) -
h_{KS}({\cal R}).
\eqlabel{er3}
\end{equation}
Now we make the reasonable conjecture that the microscopic and the macroscopic
escape-rates are equal, which leads to an expression for the diffusion
coefficient $D$ given by
\begin{equation}
D =
\lim_{L\rightarrow\infty}\frac{L^{2}}{A}\left[\sum_{\lambda_{i}>0}\lambda_{i}
({\cal R}) -
h_{KS}({\cal R})\right],
\eqlabel{er4}
\end{equation}
where we have taken the large system limit in order to remove terms of higher
order in $1/L$ resulting from deviations of the actual dynamics from the
diffusion law. In the event that the limit on the right hand side of \eq{er4}
exists, one has an expression for a macroscopic transport coefficient in terms
of microscopic dynamical quantities.  The escape-rate formalism has been
applied by Gaspard and Baras\cite{gasbar} to determine the diffusion
coefficient of a particle moving in a dense array of hard disk scatterers,
where the centers of the scatterers are placed at the vertices of a triangular
lattice, and by van Beijeren, Dorfman, and Latz, to determine the KS entropy on
the repeller of a dilute, random Lorentz gas with hard disk or hard sphere
scatterers\cite{vbld}.

\section{Largest Lyapunov exponent of a gas of hard spheres at low density}
\label{sec:lyap}

\subsection{The hard sphere gas}

Often the calculation of chaotic  characteristics of a system  can only be done
numerically. It would be preferable if one could at least find approximate
values for such quantities using more analytical methods, and thus gain some
insight  into the relevant physical processes. For the Lorentz Gas at low
densities, this was done by Dorfman, Van Beijeren and others
\cite{Dorfman2,vbld,Beijerenetal,Latz}. Here we present a calculation of the
largest Lyapunov exponent for a system that is  closer to a real gas, namely a
dilute hard sphere gas. A brief presentation of this calculation can be found
in \citeref{myself}.

We take $N$ hard spheres in a volume $V$, in $d$ dimensions.  The diameter of
the hard  spheres is $\sigma$, the reduced density is defined as the
dimensionless number $\tilde{n}=N\sigma^d/V$. We work in the thermodynamic
limit $N,V\rightarrow\infty$, keeping $\tilde{n}$ fixed (but small). In this
limit we need not be concerned with  boundary conditions, but one  may think of
periodic boundary conditions, which have been used in Molecular Dynamics
simulations to which we will eventually compare our results.

The phase-space of the hard sphere gas consists of the positions $\{\vvr_i\}$
and velocities $\{\vv_i\}$ of all $N$ particles. To calculate the largest
Lyapunov exponent, denoted here by $\lambda_+$, we need to consider two
infinitesimally close trajectories in phase-space,
$\Gamma=(\vvr_1,\vv_1,\ldots,\vvr_N,\vv_N)$ and $\Gamma+\delta\Gamma$ $=$
$\Gamma$ $+$  $(\vdr_1,\vdv_1,\ldots,\vdr_N,\vdv_N)$. The dynamics of the
$\vdr_i$ and $\vdv_i$  is found from linearizing the dynamics of $\vvr_i$ and
$\vv_i$, which consists of a sequence of free flights and binary collisions. In
free flight, there  are continuous changes,
\begin{eqnarray}
      \dot{\vvr_i} &=& \vv_i \nonumber  \\
    \dot{\vv_i} &=& 0   \nonumber \\
        \dot{\vdr_i} &=& \vdv_i  \nonumber \\
   \dot{\vdv_i} &=& 0,
\eqlabel{ll1}
\end{eqnarray}
and at collisions the values of the two particles $i$ and $j$ change
discontinuously according to:
\begin{eqnarray}
        \vvrp_j &=& \vvr_j  \nonumber \\
   \vvp_j &=& \vv_j + (\vv_{ij}\cdot\vhat{\sigma})\vhat{\sigma} \nonumber  \\
     \vdrp_j &=& \vdr_j + (\vdr_{ij}\cdot\vhat{\sigma})\vhat{\sigma}
\nonumber  \\
        \vdvp_j &=& \vdv_j + (\vdv_{ij}\cdot\vhat{\sigma})\vhat{\sigma} +
\matr{Q}
     (\vdr_i-\vdr_j).
\eqlabel{ll2}
\end{eqnarray}
Primes are used to denote values right after the collision.
$\vv_{ij}=\vv_i-\vv_j$ is the relative velocity and
$\vhat{\sigma}=(\vvr_i-\vvr_j)/\sigma$ is the collision parameter. $\matr{Q}$
is the matrix\cite{myself}
\begin{equation}
      \matr{Q} =  \frac{\left[
          (\vhat{\sigma}\cdot\vect{v}_{ij})\identity
               +\vhat{\sigma}{\vect{v}_{ij}}
       \right]\cdot\left[
          (\vhat{\sigma}\cdot\vect{v}_{ij})\identity
-\vect{v}_{ij}{\vhat{\sigma}}
       \right]
  }{\sigma(\vhat{\sigma}\cdot\vect{v}_{ij})}.
\eql{Q}
\eqlabel{ll3}
\end{equation}
The non-dotted products of vectors are dyadic products and $\identity$ is the
identity matrix.

Our approach will be based on kinetic theory. We are concerned with the
distribution of $(\vvr,\vv,\vdr,\vdv)$ as a function of time. For low
densities, the evolution of the distribution function $f$ is described by a
kinetic equation\cite{DorfmanBeijerenK}. This equation is based on the
assumption that two colliding particles are uncorrelated, so that the
probability of a collision between a particle with
$(\vvr_1,\vv_1,\vdr_1,\vdv_1)$ and one with $(\vvr_2,\vv_2,\vdr_2,\vdv_2)$ is
proportional to the product  $f(\vvr_1,\vv_1,\vdr_1,\vdv_1)$
$f(\vvr_2,\vv_2,\vdr_2,\vdv_2)$.  \footnote{Of course one also has to demand
that the particles should be a distance $\sigma$ apart.}

The  kinetic equation can, unlike the ordinary Boltzmann equation, be expanded
in powers of $1/|\ln\tilde n|$ to get the low density behavior of $f$, and thus
of $\lambda_+$. We will  take a different but roughly  equivalent approach, we
will  derive the effective dynamics of the $\vdr_i$ and $\vdv_i$  for low
densities, and use that  to write down  a  kinetic equation.

\subsection{Low density dynamics -- Clock model}

The main  characteristics of the low density  region is the typically long free
flight time $\tau$ of an individual particle between collisions, compared to
the time it would take two transparent hard spheres to cross each other. If
$v_0$ is the typical thermal velocity,  the latter is $\sigma/v_0$, while
$\tau\approx 1 / (v_0\sigma^{d-1}N/V) = \sigma/(v_0\tilde n)$. Thus $\tilde{n}$
is the small parameter.

Just before a collision at time $t$ the $\vdr_i(t)$ will be
\begin{equation}
\vdr_i(t)=\vdr_i(t_0) + \vdv_i(t_0)\tau_i\, ,
\end{equation}
if $t_0$ is the time of the previous collision and $\tau_i$ is the (large) free
flight time $t-t_0$. Suppose that initially $\vdr_i(t_0)/\sigma$ and
$\vdv_i(t_0)/v_0$ are of the same order, then just before collision,
\[
  \vdr_i = \tau_i \left[\vdv_i + {\cal O}(\tilde{n})\right].
\]
We insert this into the collision rules and neglect the terms of relative order
${\cal O}(\tilde n)$:
\begin{eqnarray}
        \vdrp_j &\approx& \tau_j \vdv_j +
      \{(\tau_i\vdv_i-\tau_j\vdv_j)\cdot\vhat{\sigma}\}\vhat{\sigma}
 \nonumber \\
    \vdvp_j &\approx&  \matr{Q} (\tau_i\vdv_i-\tau_j\vdv_j).
\eql{nodr}
\end{eqnarray}
Using \eq{Q}, we see that $\vdrp_i/\sigma$ and $\vdvp_i/v_0$ are both of order
$(\vdv_i-\vdv_j)/{v_0 \tilde n}$,  and are of the order of the ratio of the
mean free time to the time it takes to traverse a particle diameter. For a
dilute gas, this ratio is large and terms of relative order $\tilde{n}$ can be
neglected. In this way we have eliminated the $\vdr_i$ from the $\vdv_i$
dynamics.

The neglected terms were of relative order $\tilde n$, relative to either
$\tau_i\vdv_i$ or $\tau_j\vdv_j$.  These two are not necesarily of the same
order.  Now, if one of them  is one or more orders of $\tilde{n}$ higher than
the other, we should also neglect it. If they are both of the same order, we
should keep both. To know which terms to neglect, we have to keep track of the
orders of $\tilde n$ in $\vdv_i$. For that purpose, we define
\begin{equation}
       \vdv_i = v_0 (\tilde{n})^{-k_i} \vhat{e}_i,
\eql{kdef}
\end{equation}
where $\|\vhat{e}_i\|=1$. The number $k_i$ counts the number of orders of
$\tilde n$ and we will call  this the {\em clock value} of particle $i$. The
clock values are real numbers at this point, but later will be  approximated by
integers. Inserting \eq{kdef}  into \eq{nodr}, we can get the clock values
$k'_i$ and $k'_j$  after collision. Since, to leading order in density,
$\vdvp_i$ and $\vdvp_j$ differ only in sign, $k'_i=k'_j=k'$, with
\[
       k' = \frac{1}{-\ln{\tilde n}}\ln
        \|\matr{Q}(\tau_i\vdv_i-\tau_j\vdv_j)\|.
\]
Both $\tau_i$ and $\tau_j$ are typically of order $\sigma/(v_0\tilde n)$. This
means that if  $k_i>k_j$,  we should neglect the term with $\vdv_j$, and if
$k_i<k_j$,  we should neglect the other term. This yields:
\[
     k' = k_D + \frac{1}{-\ln{\tilde
         n}}\ln\|\matr{Q}\tau_D\vhat{e}_D\|,
\]
where $D=i$ if $k_i>k_j$, and $D=j$ otherwise.  Particle $D$ is called the
dominant particle. Using  the property $\tau_D={\cal O} (\sigma/(v_0\tilde n))$
and the explicit form of $\matr{Q}$ from \eq{Q}, one gets
\[
  k' = k_D + 1 + {\cal O}(\frac{1}{\ln\tilde n}).
\]
So far we have  ignored the possibility that  $k_i=k_j$.  But the resulting
correction  in fact  only contributes to the ${\cal O}(1/\ln\tilde n)$ part.

Differences  in the number of collisions suffered by different particles, cause
the clock values to not be all the same, even if they were  so initially.  They
are indispensible for determining the magnitudes of  postcollisional  velocity
deviations. But in fact  no more is needed! For  if we know how fast the clock
values grow, we know the linear growth of $\ln\|\vdv\|$, which is precisely
the largest Lyapunov exponent.  So we define the clock speed as
\[
      w = \lim_{t\rightarrow\infty} \frac{<k(t)>}{\bar{\nu} t},
\]
in which $<k>$ is the average clock value and $\bar{\nu}$ is the average
collision frequency.\footnote{In a hard sphere gas in $d$ dimensions,
$\bar\nu=\frac{2\pi^{\frac{d-1}{2}}}{\Gamma(\frac{d}{2})}
\sqrt{\frac{k_BT}{m\sigma^2}}\;\tilde{n}$. } Because we  extracted $\bar\nu$,
which is ${\cal O}(\tilde n)$, this clock speed is of order 1.

The Lyapunov exponent is related to $w$ via
\[
        \lambda_+ = - w \bar \nu \ln \tilde n .
\]
The clock speed $w$ will be calculated in an expansion in $1/|\ln\tilde n|$.
The leading order for low density, calculated in the next section, is a
non-trivial constant. This behavior of $\lambda_+$ as a function of density had
been conjectured already by Krylov\cite{Krylov}, however without  a nontrivial
prefactor $w$. A first estimate of $w$ was made by Stoddard and
Ford\cite{StoddardFord}, who got $w =\ln N$. This would mean that there is no
thermodynamic limit for the Lyapunov exponent. Some numerical
simulations\cite{Searles} have been interpreted as  supporting  the logarithmic
divergence with $N$, but with a prefactor much smaller than $1$. However, we
will find a finite thermodynamic limit for $w$ and show that, even in a
mean-field approach that fully ignores local density fluctuations, it
approaches this thermodynamic value  so slowly that in the range of particle
numbers accessible to simulations one could not distiguish between saturation
or slow but steady increase.

\subsection{Kinetic approach}

For low densities, we may describe the effective dynamics of the clock values
by
\begin{equation}
    k_i' = k_j ' = \max(k_i,k_j) + 1,
\eql{collrule}
\end{equation}
where the collision pairs  $(i,j)$  are chosen completely randomly with
Poisson distributed  collision times. The model with this dynamics we will call
the clock model.  For simplicity we will consider  integer clock values only,
though this restriction is by no means necessary or important. The clock  speed
$w $ found in this model gives the leading term in the density expansion of the
Lyapunov exponent:
\begin{equation}
      \lambda_+ = w \bar \nu \left[ - \ln \tilde n +
         {\cal O}(1)\right].
\eql{lle}
\end{equation}
A distribution function $f_k(t)$ will denote the fraction of particles having
clock value $k$ at time $t$.  From the dynamics specified above we can derive
an equation for the distribution function $f_k(t)$ of clock values.  We  expect
the clock values to grow linearly with time. If they all grow at the same rate,
we  have
\[
      f_k(t) = g(k-w\bar\nu t),
\]
with $w$ as defined before, because
\[
\lim_{t\rightarrow\infty} \frac{1}{\bar\nu t}
   \sum_{k=-\infty}^{\infty} g(k-w\bar\nu t) k
 =\lim_{t\rightarrow\infty} \frac{1}{\bar\nu t}
   \sum_{x=-\infty}^{\infty} g(x) (x+w\bar\nu t) = w.
\]
So once we have a  kinetic equation for $f_k(t)$, we will look for these  {\em
propagating solutions}.

Consider the contributions to $\dd{f_k}{t}$ from collisions in which a clock
value $k$ is lost. In each collision where a particle with $k$ enters, the $k$
gets lost, so the fraction of particles with $k$ decreases  at a rate $\bar \nu
f_k(t)$ due to these processes. There are also processes which increase the
fraction of particles having $k$. For  these the  larger incoming clock value
should be $k-1$. We have to distiguish collisions with equal incoming clock
values, both $k-1$, from ones with  different incoming clock values $k-1$ and
$l$.  The rate for  the latter type of collisions is $\bar\nu f_{k-1}
\sum_{l=-\infty}^{k-2} f_l$.   For the former   ones the rate is only $1/2$
$\bar \nu f^2_{k-1}$, because the two particles are drawn from the same
fraction $f_{k-1}$, and it doesn't matter in which order they are picked. In
either case the number of particles with clock value $k$ increases by two,  so
we get the  kinetic equation:
\[
   \dd{f_k(t)}{\bar\nu t} = - f_k(t) + f^2 _{k-1} + 2
     f_{k-1}\sum_{l=-\infty}^{k-2} f_l .
\]
We scale out $\bar\nu$ by defining a new time variable $\tau=\bar\nu t$. The
equation is  simplified further by replacing the fraction $f_k$ of particles
having clock value $k$, by the fraction $C_k$ of particles that have clock
values $k$ {\em or less}:
\[
     C_k(\tau) = \sum_{l=-\infty}^{k} f_l(\tau).
\]
The  kinetic equation then takes the short form:
\begin{equation}
   \dd{C_k}{\tau} = - C_k+ C^2_{k-1}.
       \eql{BE}
\end{equation}

\subsection{Front propagation}

Let us investigate the solutions of \eq{BE}. We see that,  given $C_{k-1}$ and
the initial value of $C_k$, we can solve for $C_k$:
\[
     C_k(\tau) = e^{-\tau} \left[
    C_k(0)+\int_0^{\tau} e^{\tau'} C^2_{k-1}(\tau')d\tau'\right].
\]
If $C_{k}=0$ initially, it will remain zero, because $0\leq C_{k-1} \leq C_k$.
We take initial conditions such that $C_{k<1}(0) = 0$ and $C_{k>0}=1$, i.e. all
particles have $k=1$. The solutions $C_k$ are all polynomials in $\exp(-\tau)$,
but with increasing $k$ the order grows exponentially.  Nonetheless, we
calculated the $C_k$ up to $k=32$ using a computer program to handle the
analytic manipulations.  The $C_k$'s are plotted in \fig{anasol} for
$\tau=0,2,4,6$ and $10$.

\begin{figure}[h]
 \centerline{\psfig{figure=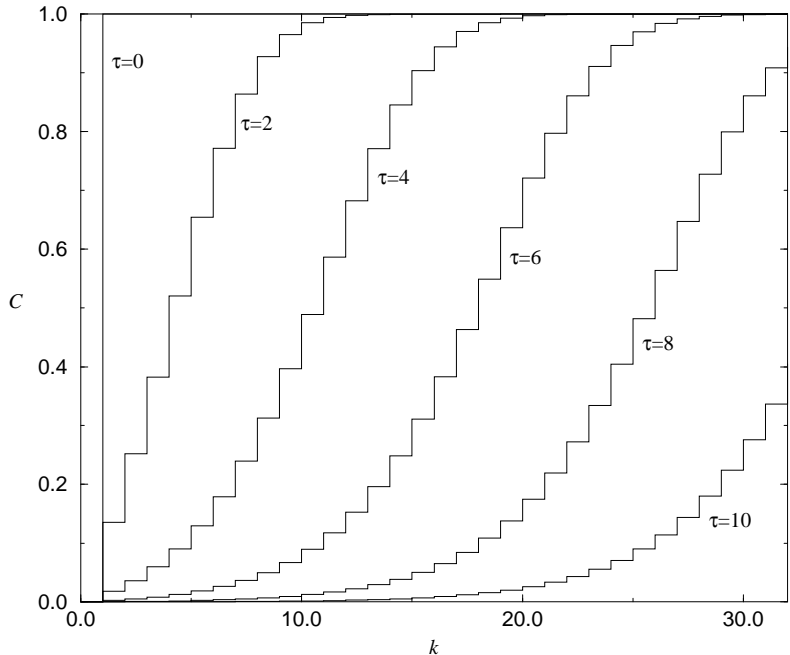,width=\textwidth,angle=-90}}
 \caption{
Time evolution of the clock
value
distribution.
All initial clock values are $0$.
}
        \figl{anasol}
\end{figure}

We see in \fig{anasol} that the initial distribution changes to some smoother
shape, and moves to the right.  After a while the shape seems to stay constant.
We can now view \eq{BE} as  describing a propagating front: $C\equiv 0$ is a
stable phase and $C\equiv 1$ is an unstable phase. On the left we have the
stable phase, on the right, the unstable phase and in between is the
intermediate region called {\em the front }, that propagates to the right, into
the unstable phase. The velocity at which it moves to the right is $w$. From
\fig{anasol} we see that for $\tau=10$, the speed is still increasing, and is
about 3.8.

Front propagation into an unstable phase comes in two
flavors\cite{VanSaarloos3}. In general, the instability of the unstable phase
sets a velocity $w^*$ by which small perturbations propagate to the right. This
velocity $w^*$ is determined from the linearized  equation describing the front
propagation around the unstable phase. For {\em pulled} fronts $w^*$ is the
asymptotic velocity of any solution with an initial shape that is sufficiently
steep. If, however, no solution of the full non-linear front equation with this
velocity exists, or if it is unstable with respect to some nonlinear
perturbation, the velocity is set by these non-linearities and the front is
called {\em pushed}. In that case the velocity is higher than $w^*$. We will
assume that in  our case the front is {\em pulled}.  As we do not know of a
general criterion for deciding whether a front is pushed  or pulled, we will
use the results of computer simulations for the validation of our  assumption.

We want to find a solution to \eq{BE} of the form
\[
      C_k(\tau) = F(k-w\tau) = F(x).
\]
which means
\begin{equation}
      -w \ddh{F(x)}{x} = - F(x) + F^2(x-1).
    \eql{dde}
\end{equation}
Now $C$ is a positive, increasing function of $k$, bounded between $0$ and $1$.
The function $F$ should have all these properties too. So we are looking for a
solution of a form such as depicted in \fig{front}.

\begin{figure}[h]
 \centerline{\psfig{figure=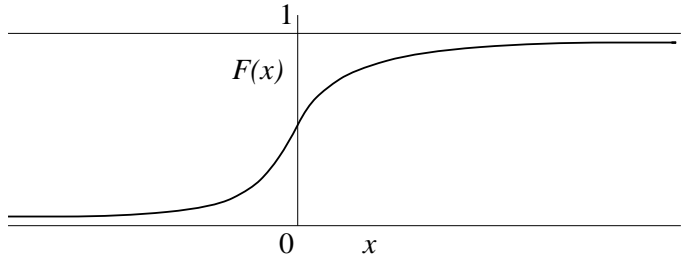,width=9cm,angle=-90}}
   \caption{   Clock value distribution as a propagating front}
      \figl{front}
\end{figure}

For a pulled front, we have to investigate the  \eq{dde} linearized around the
unstable phase.  Writing $F=1-\Delta$, we get
\begin{equation}
     - w \ddh{\Delta(x)}{x} = - \Delta(x) + 2 \Delta(x-1) + {\cal
   O}(\Delta^2).
   \eql{lin}
\end{equation}
Given $w$, the  asymptotic solution of \eq{lin} is given by a sum of
exponentials\cite{Bellman}
\begin{equation}
 \Delta (x) = \sum_i A_i e^{-\gamma_i x}.
\eql{exponentials}
\end{equation}
The possible values of $\gamma_i$ are found by inserting $\exp(-\gamma x)$ into
the linearized equation (in case of degeneracies, $A_i$ should be replaced by a
polynomial in $x$). This gives $w$ as a function of $\gamma$:
\begin{equation}
  w(\gamma) = (2e^\gamma-1)/\gamma.
\eql{disp}
\end{equation}
The $\gamma_i$  may be complex, and, for given $w$,  there are infinitely many
of them. However, we know that $\Delta$ should be  monotonic, so the most
slowly decaying term in the sum should have a real positive $\gamma$.

{}From the plot of \eq{disp} in \fig{disp}, we see that there are three cases:
If $w$ is larger than some critical $w^*$, there are two $\gamma$'s, which are
real and positive. If $w=w^*$ these two become degenerate and for $w<w^*$ they
become complex. But for the slowest term to have complex $\gamma$ was not
allowed by the condition of monotonicity of $F$: asymptotically the function
would oscillate.  So we conclude that, for the the Ansatz of a propagating
front to work, the velocity should be at least $w^*$, the minimal $w$ from
\eq{disp} for positive real $\gamma$. This value can be expressed in terms of
Lambert's $W$ function:\footnote{Defined as $W(x)\;\exp[W(x)] = x$, where the
branch analytic in $0$ is meant.}
\[
 w^* = \frac{-1}{W(\frac{-1}{2e})} \approx 4.31107\ldots,
\]

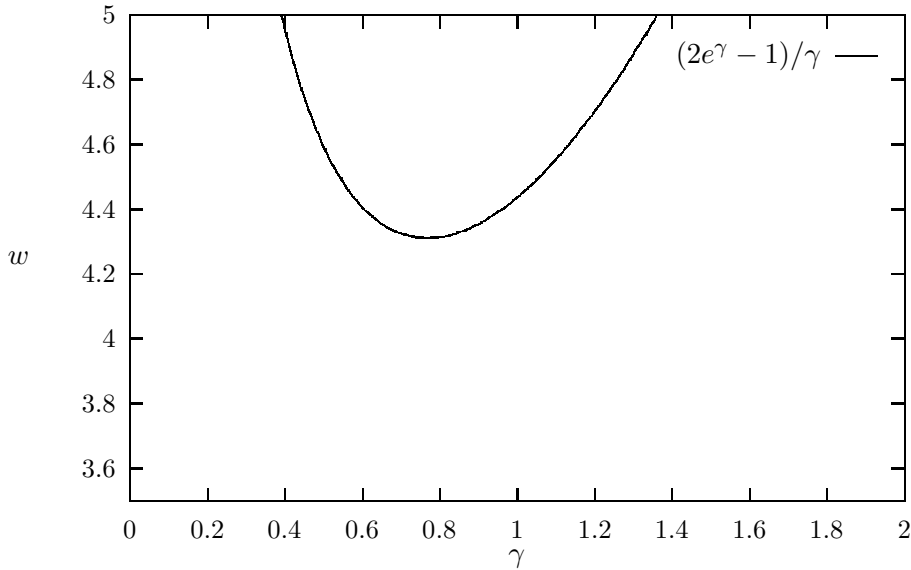
\begin{figure}[h]
 \centerline{
% GNUPLOT: LaTeX picture
\setlength{\unitlength}{0.240900pt}
\ifx\plotpoint\undefined\newsavebox{\plotpoint}\fi
\begin{picture}(1500,900)(0,0)
\font\gnuplot=cmr10 at 10pt
\gnuplot
\sbox{\plotpoint}{\rule[-0.200pt]{0.400pt}{0.400pt}}%
\put(220.0,113.0){\rule[-0.200pt]{0.400pt}{184.048pt}}
\put(220.0,164.0){\rule[-0.200pt]{4.818pt}{0.400pt}}
\put(198,164){\makebox(0,0)[r]{3.6}}
\put(1416.0,164.0){\rule[-0.200pt]{4.818pt}{0.400pt}}
\put(220.0,266.0){\rule[-0.200pt]{4.818pt}{0.400pt}}
\put(198,266){\makebox(0,0)[r]{3.8}}
\put(1416.0,266.0){\rule[-0.200pt]{4.818pt}{0.400pt}}
\put(220.0,368.0){\rule[-0.200pt]{4.818pt}{0.400pt}}
\put(198,368){\makebox(0,0)[r]{4}}
\put(1416.0,368.0){\rule[-0.200pt]{4.818pt}{0.400pt}}
\put(220.0,470.0){\rule[-0.200pt]{4.818pt}{0.400pt}}
\put(198,470){\makebox(0,0)[r]{4.2}}
\put(1416.0,470.0){\rule[-0.200pt]{4.818pt}{0.400pt}}
\put(220.0,571.0){\rule[-0.200pt]{4.818pt}{0.400pt}}
\put(198,571){\makebox(0,0)[r]{4.4}}
\put(1416.0,571.0){\rule[-0.200pt]{4.818pt}{0.400pt}}
\put(220.0,673.0){\rule[-0.200pt]{4.818pt}{0.400pt}}
\put(198,673){\makebox(0,0)[r]{4.6}}
\put(1416.0,673.0){\rule[-0.200pt]{4.818pt}{0.400pt}}
\put(220.0,775.0){\rule[-0.200pt]{4.818pt}{0.400pt}}
\put(198,775){\makebox(0,0)[r]{4.8}}
\put(1416.0,775.0){\rule[-0.200pt]{4.818pt}{0.400pt}}
\put(220.0,877.0){\rule[-0.200pt]{4.818pt}{0.400pt}}
\put(198,877){\makebox(0,0)[r]{5}}
\put(1416.0,877.0){\rule[-0.200pt]{4.818pt}{0.400pt}}
\put(220.0,113.0){\rule[-0.200pt]{0.400pt}{4.818pt}}
\put(220,68){\makebox(0,0){0}}
\put(220.0,857.0){\rule[-0.200pt]{0.400pt}{4.818pt}}
\put(342.0,113.0){\rule[-0.200pt]{0.400pt}{4.818pt}}
\put(342,68){\makebox(0,0){0.2}}
\put(342.0,857.0){\rule[-0.200pt]{0.400pt}{4.818pt}}
\put(463.0,113.0){\rule[-0.200pt]{0.400pt}{4.818pt}}
\put(463,68){\makebox(0,0){0.4}}
\put(463.0,857.0){\rule[-0.200pt]{0.400pt}{4.818pt}}
\put(585.0,113.0){\rule[-0.200pt]{0.400pt}{4.818pt}}
\put(585,68){\makebox(0,0){0.6}}
\put(585.0,857.0){\rule[-0.200pt]{0.400pt}{4.818pt}}
\put(706.0,113.0){\rule[-0.200pt]{0.400pt}{4.818pt}}
\put(706,68){\makebox(0,0){0.8}}
\put(706.0,857.0){\rule[-0.200pt]{0.400pt}{4.818pt}}
\put(828.0,113.0){\rule[-0.200pt]{0.400pt}{4.818pt}}
\put(828,68){\makebox(0,0){1}}
\put(828.0,857.0){\rule[-0.200pt]{0.400pt}{4.818pt}}
\put(950.0,113.0){\rule[-0.200pt]{0.400pt}{4.818pt}}
\put(950,68){\makebox(0,0){1.2}}
\put(950.0,857.0){\rule[-0.200pt]{0.400pt}{4.818pt}}
\put(1071.0,113.0){\rule[-0.200pt]{0.400pt}{4.818pt}}
\put(1071,68){\makebox(0,0){1.4}}
\put(1071.0,857.0){\rule[-0.200pt]{0.400pt}{4.818pt}}
\put(1193.0,113.0){\rule[-0.200pt]{0.400pt}{4.818pt}}
\put(1193,68){\makebox(0,0){1.6}}
\put(1193.0,857.0){\rule[-0.200pt]{0.400pt}{4.818pt}}
\put(1314.0,113.0){\rule[-0.200pt]{0.400pt}{4.818pt}}
\put(1314,68){\makebox(0,0){1.8}}
\put(1314.0,857.0){\rule[-0.200pt]{0.400pt}{4.818pt}}
\put(1436.0,113.0){\rule[-0.200pt]{0.400pt}{4.818pt}}
\put(1436,68){\makebox(0,0){2}}
\put(1436.0,857.0){\rule[-0.200pt]{0.400pt}{4.818pt}}
\put(220.0,113.0){\rule[-0.200pt]{292.934pt}{0.400pt}}
\put(1436.0,113.0){\rule[-0.200pt]{0.400pt}{184.048pt}}
\put(220.0,877.0){\rule[-0.200pt]{292.934pt}{0.400pt}}
\put(45,495){\makebox(0,0){$w$}}
\put(828,23){\makebox(0,0){$\gamma$}}
\put(220.0,113.0){\rule[-0.200pt]{0.400pt}{184.048pt}}
\put(1306,812){\makebox(0,0)[r]{$(2e^{\gamma}-1)/\gamma$}}
\multiput(458.59,870.15)(0.488,-2.013){13}{\rule{0.117pt}{1.650pt}}
\multiput(457.17,873.58)(8.000,-27.575){2}{\rule{0.400pt}{0.825pt}}
\multiput(466.58,839.22)(0.492,-1.961){21}{\rule{0.119pt}{1.633pt}}
\multiput(465.17,842.61)(12.000,-42.610){2}{\rule{0.400pt}{0.817pt}}
\multiput(478.58,793.91)(0.492,-1.746){21}{\rule{0.119pt}{1.467pt}}
\multiput(477.17,796.96)(12.000,-37.956){2}{\rule{0.400pt}{0.733pt}}
\multiput(490.58,753.99)(0.493,-1.408){23}{\rule{0.119pt}{1.208pt}}
\multiput(489.17,756.49)(13.000,-33.493){2}{\rule{0.400pt}{0.604pt}}
\multiput(503.58,718.30)(0.492,-1.315){21}{\rule{0.119pt}{1.133pt}}
\multiput(502.17,720.65)(12.000,-28.648){2}{\rule{0.400pt}{0.567pt}}
\multiput(515.58,687.71)(0.492,-1.186){21}{\rule{0.119pt}{1.033pt}}
\multiput(514.17,689.86)(12.000,-25.855){2}{\rule{0.400pt}{0.517pt}}
\multiput(527.58,660.26)(0.492,-1.013){21}{\rule{0.119pt}{0.900pt}}
\multiput(526.17,662.13)(12.000,-22.132){2}{\rule{0.400pt}{0.450pt}}
\multiput(539.58,636.90)(0.493,-0.814){23}{\rule{0.119pt}{0.746pt}}
\multiput(538.17,638.45)(13.000,-19.451){2}{\rule{0.400pt}{0.373pt}}
\multiput(552.58,615.96)(0.492,-0.798){21}{\rule{0.119pt}{0.733pt}}
\multiput(551.17,617.48)(12.000,-17.478){2}{\rule{0.400pt}{0.367pt}}
\multiput(564.58,597.51)(0.492,-0.625){21}{\rule{0.119pt}{0.600pt}}
\multiput(563.17,598.75)(12.000,-13.755){2}{\rule{0.400pt}{0.300pt}}
\multiput(576.58,582.65)(0.492,-0.582){21}{\rule{0.119pt}{0.567pt}}
\multiput(575.17,583.82)(12.000,-12.824){2}{\rule{0.400pt}{0.283pt}}
\multiput(588.00,569.92)(0.590,-0.492){19}{\rule{0.573pt}{0.118pt}}
\multiput(588.00,570.17)(11.811,-11.000){2}{\rule{0.286pt}{0.400pt}}
\multiput(601.00,558.92)(0.600,-0.491){17}{\rule{0.580pt}{0.118pt}}
\multiput(601.00,559.17)(10.796,-10.000){2}{\rule{0.290pt}{0.400pt}}
\multiput(613.00,548.93)(0.758,-0.488){13}{\rule{0.700pt}{0.117pt}}
\multiput(613.00,549.17)(10.547,-8.000){2}{\rule{0.350pt}{0.400pt}}
\multiput(625.00,540.93)(1.123,-0.482){9}{\rule{0.967pt}{0.116pt}}
\multiput(625.00,541.17)(10.994,-6.000){2}{\rule{0.483pt}{0.400pt}}
\multiput(638.00,534.94)(1.651,-0.468){5}{\rule{1.300pt}{0.113pt}}
\multiput(638.00,535.17)(9.302,-4.000){2}{\rule{0.650pt}{0.400pt}}
\multiput(650.00,530.95)(2.472,-0.447){3}{\rule{1.700pt}{0.108pt}}
\multiput(650.00,531.17)(8.472,-3.000){2}{\rule{0.850pt}{0.400pt}}
\put(662,527.17){\rule{2.500pt}{0.400pt}}
\multiput(662.00,528.17)(6.811,-2.000){2}{\rule{1.250pt}{0.400pt}}
\put(674,525.67){\rule{3.132pt}{0.400pt}}
\multiput(674.00,526.17)(6.500,-1.000){2}{\rule{1.566pt}{0.400pt}}
\put(687,525.67){\rule{2.891pt}{0.400pt}}
\multiput(687.00,525.17)(6.000,1.000){2}{\rule{1.445pt}{0.400pt}}
\put(699,526.67){\rule{2.891pt}{0.400pt}}
\multiput(699.00,526.17)(6.000,1.000){2}{\rule{1.445pt}{0.400pt}}
\multiput(711.00,528.61)(2.695,0.447){3}{\rule{1.833pt}{0.108pt}}
\multiput(711.00,527.17)(9.195,3.000){2}{\rule{0.917pt}{0.400pt}}
\multiput(724.00,531.60)(1.651,0.468){5}{\rule{1.300pt}{0.113pt}}
\multiput(724.00,530.17)(9.302,4.000){2}{\rule{0.650pt}{0.400pt}}
\multiput(736.00,535.60)(1.651,0.468){5}{\rule{1.300pt}{0.113pt}}
\multiput(736.00,534.17)(9.302,4.000){2}{\rule{0.650pt}{0.400pt}}
\multiput(748.00,539.59)(1.033,0.482){9}{\rule{0.900pt}{0.116pt}}
\multiput(748.00,538.17)(10.132,6.000){2}{\rule{0.450pt}{0.400pt}}
\multiput(760.00,545.59)(1.123,0.482){9}{\rule{0.967pt}{0.116pt}}
\multiput(760.00,544.17)(10.994,6.000){2}{\rule{0.483pt}{0.400pt}}
\multiput(773.00,551.59)(0.758,0.488){13}{\rule{0.700pt}{0.117pt}}
\multiput(773.00,550.17)(10.547,8.000){2}{\rule{0.350pt}{0.400pt}}
\multiput(785.00,559.59)(0.758,0.488){13}{\rule{0.700pt}{0.117pt}}
\multiput(785.00,558.17)(10.547,8.000){2}{\rule{0.350pt}{0.400pt}}
\multiput(797.00,567.59)(0.824,0.488){13}{\rule{0.750pt}{0.117pt}}
\multiput(797.00,566.17)(11.443,8.000){2}{\rule{0.375pt}{0.400pt}}
\multiput(810.00,575.58)(0.600,0.491){17}{\rule{0.580pt}{0.118pt}}
\multiput(810.00,574.17)(10.796,10.000){2}{\rule{0.290pt}{0.400pt}}
\multiput(822.00,585.58)(0.600,0.491){17}{\rule{0.580pt}{0.118pt}}
\multiput(822.00,584.17)(10.796,10.000){2}{\rule{0.290pt}{0.400pt}}
\multiput(834.00,595.58)(0.543,0.492){19}{\rule{0.536pt}{0.118pt}}
\multiput(834.00,594.17)(10.887,11.000){2}{\rule{0.268pt}{0.400pt}}
\multiput(846.00,606.58)(0.539,0.492){21}{\rule{0.533pt}{0.119pt}}
\multiput(846.00,605.17)(11.893,12.000){2}{\rule{0.267pt}{0.400pt}}
\multiput(859.00,618.58)(0.496,0.492){21}{\rule{0.500pt}{0.119pt}}
\multiput(859.00,617.17)(10.962,12.000){2}{\rule{0.250pt}{0.400pt}}
\multiput(871.58,630.00)(0.492,0.539){21}{\rule{0.119pt}{0.533pt}}
\multiput(870.17,630.00)(12.000,11.893){2}{\rule{0.400pt}{0.267pt}}
\multiput(883.58,643.00)(0.493,0.536){23}{\rule{0.119pt}{0.531pt}}
\multiput(882.17,643.00)(13.000,12.898){2}{\rule{0.400pt}{0.265pt}}
\multiput(896.58,657.00)(0.492,0.582){21}{\rule{0.119pt}{0.567pt}}
\multiput(895.17,657.00)(12.000,12.824){2}{\rule{0.400pt}{0.283pt}}
\multiput(908.58,671.00)(0.492,0.625){21}{\rule{0.119pt}{0.600pt}}
\multiput(907.17,671.00)(12.000,13.755){2}{\rule{0.400pt}{0.300pt}}
\multiput(920.58,686.00)(0.492,0.669){21}{\rule{0.119pt}{0.633pt}}
\multiput(919.17,686.00)(12.000,14.685){2}{\rule{0.400pt}{0.317pt}}
\multiput(932.58,702.00)(0.493,0.616){23}{\rule{0.119pt}{0.592pt}}
\multiput(931.17,702.00)(13.000,14.771){2}{\rule{0.400pt}{0.296pt}}
\multiput(945.58,718.00)(0.492,0.669){21}{\rule{0.119pt}{0.633pt}}
\multiput(944.17,718.00)(12.000,14.685){2}{\rule{0.400pt}{0.317pt}}
\multiput(957.58,734.00)(0.492,0.755){21}{\rule{0.119pt}{0.700pt}}
\multiput(956.17,734.00)(12.000,16.547){2}{\rule{0.400pt}{0.350pt}}
\multiput(969.58,752.00)(0.493,0.695){23}{\rule{0.119pt}{0.654pt}}
\multiput(968.17,752.00)(13.000,16.643){2}{\rule{0.400pt}{0.327pt}}
\multiput(982.58,770.00)(0.492,0.755){21}{\rule{0.119pt}{0.700pt}}
\multiput(981.17,770.00)(12.000,16.547){2}{\rule{0.400pt}{0.350pt}}
\multiput(994.58,788.00)(0.492,0.798){21}{\rule{0.119pt}{0.733pt}}
\multiput(993.17,788.00)(12.000,17.478){2}{\rule{0.400pt}{0.367pt}}
\multiput(1006.58,807.00)(0.492,0.841){21}{\rule{0.119pt}{0.767pt}}
\multiput(1005.17,807.00)(12.000,18.409){2}{\rule{0.400pt}{0.383pt}}
\multiput(1018.58,827.00)(0.493,0.774){23}{\rule{0.119pt}{0.715pt}}
\multiput(1017.17,827.00)(13.000,18.515){2}{\rule{0.400pt}{0.358pt}}
\multiput(1031.58,847.00)(0.492,0.841){21}{\rule{0.119pt}{0.767pt}}
\multiput(1030.17,847.00)(12.000,18.409){2}{\rule{0.400pt}{0.383pt}}
\multiput(1043.59,867.00)(0.477,1.044){7}{\rule{0.115pt}{0.900pt}}
\multiput(1042.17,867.00)(5.000,8.132){2}{\rule{0.400pt}{0.450pt}}
\put(1328.0,812.0){\rule[-0.200pt]{15.899pt}{0.400pt}}
\end{picture}
}
        \caption{
Velocity versus asymptotic decay rate $\gamma$.}
     \figl{disp}
\end{figure}

\noindent and it is the velocity set by the instability that we mentioned
before. So it {\em is} the velocity $w $ we were after.  Its value is
compatible with the estimate 3.8 from \fig{anasol} (a lower value than 3.8
would  not),  which supports the assumption of a pulled front.  This concludes
the calculation of the leading order in \eq{lle} of the largest Lyapunov
exponent  in the infinite system limit within the framework of our clock model,
but we still have to consider the effects of a finite number of particles.

\subsection{Large finite $N$ effects}

The kinetic equation works  well in the thermodynamic limit, but to compare our
results with simulations we need to correct for the  effects of the finiteness
of the number of  particles. The clock model is very suitable for  simple
simulations. We take a set of $N$ integer numbers $\{k_i\}$. In each time step
two of them are picked at random, and the collision rule in \eq{collrule} is
applied to them. We do this $T$ times. The average number of collisions per
particle is then $2T/N$, because in each of the $T$ collisions two particles
are involved.  An estimate for the clock speed is the average clock value
$\sum_i k_i / N$ divided by the average number of collisions per particle. For
large $T$ this approaches the clock speed $w_N$, so
\[
       w_N = \lim_{T\rightarrow\infty} \frac{1}{2T}\sum_{i=1}^N k_i.
\]
This simulation is not even a bad chacterization of what happens with the clock
values in the hard sphere gas, because at low densities there is very little
correlation between pairs of particles involved  in subsequent collisions. It
is  very similar to the  variant of the Direct Simulation Monte Carlo method
used by Dellago and Posch\cite{Dellago3} for the calculation of Lyapunov
exponents in a ``spatially homogeneous system".

Clock speeds for different numbers of particles are plotted in \fig{finiteN}.
Each point is determined from a single run with 2000 collisions per particle.
The sums of all clock values at 20 times were fitted to a linear function. The
  slope gave $w_N$, the error in the slope gave the error in $w_N$. The errors
are always less than 0.5 percent. One sees that $w_N$  increases slowly with
increasing $N$. It is possible that it saturates at the value of $4.311$, but
this  cannot yet be seen even for half a million particles.

A natural thought is that the $N$ dependence of $w$ is due to correlations
between subsequent collisions: if two particles collide that already collided
just before, or that had their clock values reset shortly  before by particles
that were roughly synchronized already, the gain in clock  value of the
"slower" particle will be less than average. Hence the average  clock  speed is
reduced. However, as we will see, the reduction in clock speed can be explained
entirely on the basis of the linearized equation plus some simple  bounds on
its region of validity and this explanation does not seem to require  any
effects of correlated collisions.

In the head of the distribution, the finiteness of the number of particles
becomes important when $f_k={\cal O}(1/N)$. For this     reason, Brunet and
Derrida\cite{Brunet} treat the finite $N$ effects by the introduction of a
cutoff $\varepsilon=1/N$ in the equation, i.e. they modify the equation as soon
as $f_k<\varepsilon$. They  distinguish  three regions: one where the
non-linear behavior is dominant, one where the linear equation holds, and one
where the cutoff is effective. By glueing the solutions in these regions
together, one obtains a new $\varepsilon$-dependent front velocity.

The introduction of a cutoff in something that is supposed to be a distribution
function, which is an average over realizations, seems hard to justify. A more
satisfactory way to obtain it, is to shift all clockvalues in each realization
such that the particle with the largest clock value has $x=0$, and then average
the distribution function (this was also suggested by Kessler {\it et
al}\cite{Kessler}). Then the cutoff occurs naturally.

For $x<0$ there is a region where we can use the linearized equation. We
consider the leading term in \eq{exponentials}:
\[
        \Delta \sim \frac{c}{N}e^{-\gamma_0 x}.
\]
The prefactor $c/N$ is obtained from the fact that $\Delta={\cal O}(1/N)$ at
$x=0$ ($c$ is order $1$). The linear regime ends when $\Delta$ becomes of order
$1$, so for $x=-\ln(\frac{N}{c})/\gamma_0$. This is illustrated in
\fig{cutoff}.

\begin{figure}[tb]
 \centerline{\psfig{figure=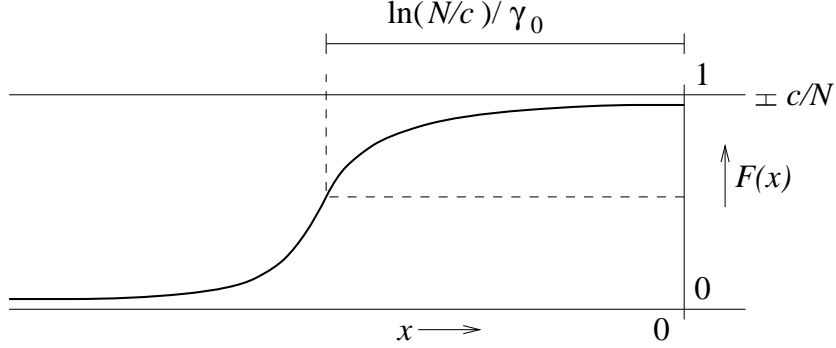,width=11cm,angle=-90}}
\caption{The cutoff at the head of the distribution and the
linear regime.}
\figl{cutoff}
\end{figure}

In contrast with the infinite system, now we only have to demand monotonicity
and positivity from the solution of the linearized equation in this interval
of width  $\ln(\frac{N}{c})/\gamma_0$.  This means that a small imaginary value
is allowed for the $\gamma$ in \eq{exponentials} with the smallest real part.
We denote $\gamma=\Reg+i\Img$. The consequential oscillations,
\[
        \Delta = a_0\, e^{-\Reg x} \cos(\Img x+\phi),
\]
should not cause sign changes in the function or its derivative. The derivative
can be sign definite if at most half a wavelength fits in the interval, i.e. if
at most
\[
 \Img=  \frac{\Reg\pi}{\ln(N/c)}.
\]
For the leading behavior $c$ can be set to $1$. Positivity of $\Delta$ itself
also poses an additional bound on $\Img$, but this can be shown to be a higher
order effect.

\begin{figure}[th]
 \centerline{\psfig{figure=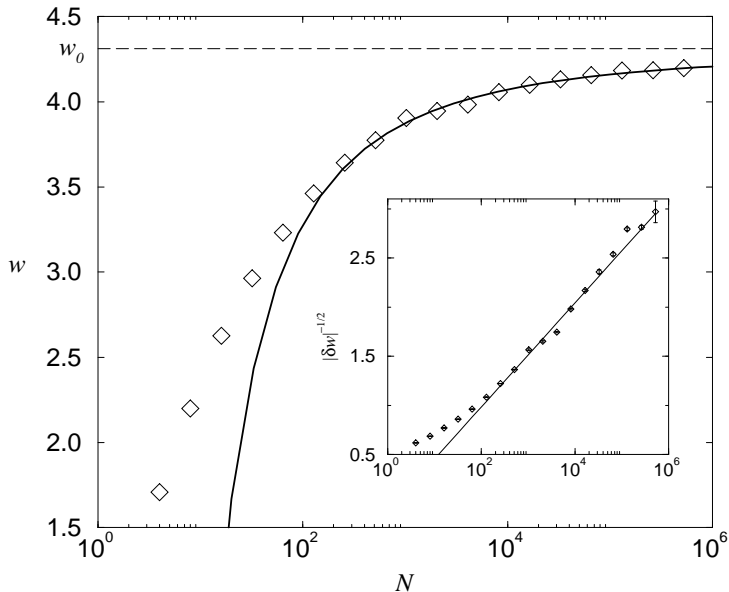,width=11cm,angle=-90}}
        \caption{Effect of finite $N$ on the clock speed. Diamonds
                are the simulations; the solid                line is the
prediction \eq{predN};
         the dashed line is the asymptotic value $4.311\ldots$.
        The inset shows the linear behavior of $1/\sqrt{w_0-w_N}$ (plotted
        with error bars) as a function of $\ln N$.
        The solid
        line in both is
        the fit to the form in \eq{form}.
          }
       \figl{finiteN}
\end{figure}

We expand $w(\gamma)$ around its minimum at $\gamma_0$:
\[
        w(\gamma)= w(\gamma_0+\dg)
                \approx w(\gamma_0)
                + \frac{1}{2} w''\dg^2
                + \frac{1}{6} w'''\dg^3
                + \ho \mbox{(higher orders)},
\]
where $w''=\ddh{^2w}{\gamma^2}(\gamma_0)$ and
$w'''=\ddh{^3w}{\gamma^3}(\gamma_0)$.    $w$ is still real, so the imaginary
part gives
\[
        0 = \Imdg \left\{
                w''\Redg
                +\frac{1}{6}w'''[3\Redg^2-\Imdg^2]
                + \ho
                \right\},
\]
where $\dg=\Redg+i\Imdg$. To first order this says that $w'' \Redg =
\frac{1}{6}w'''\Imdg^2$:  the shift in the real part of $\gamma$ is higher
order compared to the imaginary part. The new velocity is written as
$w=w_0+\dw$, and $\dw$ is obtained from the real part of equation \eq{disp}
expanded to first order:
\begin{eqnarray}
        \dw &=& \frac{1}{2}w''\left\{\Redg^2 -
                  \Imdg^2\right\}
                + \frac{1}{6} w'''\left\{ \Redg^3 -
                        \Redg\Imdg^2\right\}
                + \ho \nonumber\\
                &=& -\frac{1}{2}w''\Imdg^2 + \ho
                = -\frac{w''\pi^2\gamma_0^2}{2\ln^2(N/c)} + \ho ,
\end{eqnarray}
which coincides with Brunet and Derrida's result (when $c$ is set to $1$).
They however needed to consider how the linear region connects to the others,
while we only need that $\Delta$ is of order $1$ at the border with the
non-linear region. {}From \eq{disp}, one can show that $w''\gamma_0^2 =
(w_0-1)$, so we find the result
\begin{equation}
    w_N = w_0 - \frac{(w_0-1)\pi^2}{2\ln^2(N/c)}.
 \eql{predN}
\end{equation}

In \fig{finiteN} the simulation results are plotted together with a fit. We
calculate $\dw^{-1/2}=1/\sqrt{w_0-w_N}$, with $w_N$ taken from the simulations,
which should be a linear function of $\ln N$ for large $N$:
\begin{equation}
        \dw^{-1/2} \stackrel{N\rightarrow\infty}{\longrightarrow} a+b\ln N.
\eql{form}
\end{equation}
Indeed this behavior is seen in the inset of \fig{finiteN}. According to
\eq{predN}, $a=-\ln c$ and $b=\pi^{-1}\sqrt{2/(w_0-1)}=0.247\ldots$. The fit to
the data\footnote{As the prediction is for large $N$, only points for $N>256$
are used in the fit.} shown in the inset yields $b = 0.23 \pm 0.01$, consistent
with the theoretical prediction\footnote{By allowing changes in $a$ and $b$
simultaneously, one finds an appreciably larger range of acceptable values for
$b$ than the error in $b$  indicates.}, and $a = -0.07 \pm 0.07$. The  value of
$c$ corresponding to this $a$ is $c\approx 1.07$, so it is of order $1$ as
expected.

The values from our simulations have been compared in \citeref{myself} to
molecular dynamics simulations of hard spheres, from which $w$ is found from a
fit of $\lambda_+$ to the form \eq{lle}. The values agreed very well with the
simulations.

\subsection{Further refinements}

Comparing the results from high precision simulations of the clock model on
the  one hand and of hard disk systems with exactly the same number of
particles  and  equal collision frequency on the other hand, one finds that the
dimensionless  clock speed $w$ in the latter is significantly higher than in
the former; for a system of  $10,000$ particles the clock model gives a $w$ of
4.05 $\pm$ 0.01, whereas the corresponding hard disk value is  4.47 $\pm$ 0.02.
The cause of this difference is the velocity dependence of the collision
frequency, which is an increasing function of speed. As a result particles in
the head of the distribution tend to have a higher speed than average (a higher
 collision frequency enhances the clock value), resulting  in a clock speed
that is systematically higher than it would be in the case of a velocity
independent collision frequency. The clock model does  have a velocity
independent collision frequency indeed, which explains the  difference in $w$
between this model and the hard disk system.

An explicit calculation accounting for these effects and producing improved
estimates for $w$ in hard disk systems will appear soon\cite{betterw}.

\section{Kolmogorov-Sinai entropy of  a gas of hard spheres at low
density}
\label{sec:hKS}

In the previous section we reviewed the calculation by  Van Zon and  Van
Beijeren of the largest Lyapunov exponent for a gas of hard disks at low
densities. Here we will outline a related, but somewhat simpler calculation of
the KS entropy of a gas of hard disks or of hard spheres at low density. This
calculation, while not rigorous in a  mathematical sense, is a strong
indication of the chaotic behavior of such gases, and indicates that the
chaoticity of a dilute hard disk or hard sphere gas persists  in the
thermodynamic limit.

The starting point is the same as in the previous section.  It is important to
note that the dynamics of a hard sphere system is exactly described as free
motion of the particles, punctuated by instantaneous, binary collisions between
some pair of particles. To describe this motion and the quantities we need for
the KS-entropy, we consider the dynamical behavior of the positions and
coordinates of all the particles in the gas,
$(\vvr_1,\vv_1,\vvr_2,\vv_2,\ldots,\vvr_N,\vv_N)$, as well as a set of
deviation vectors which describe the motion of a pencil  of nearby trajectories
in phase-space, $(\vdr_1,\vdv_1,\ldots,\vdr_N,\vdv_N)$. The  equations of
motion for these quantities between the binary collisions are given by \eq{ll1}
and the changes of these quantities at a collision between particles $i$ and
$j$ are given in Eqs.~(\ref{eq:ll2},\ref{eq:ll3}).

To proceed with the calculation of the KS-entropy, we will suppose that the
Pesin formula holds, i.e., that the KS-entropy is the sum of the positive
Lyapunov exponents.  This sum can be obtained by considering the growth in time
of the volume of the $dN$-dimensional projection of an arbitrary, infinitesimal
volume in the full $2dN$-dimensional phase space.  This observation requires
some explanation, as follows. The typical rate of separation of two arbitrary,
but infinitesimally close, points in phase-space, for a hyperbolic  system,
will be exponential and the rate will be given by the largest Lyapunov
exponent. If we consider a typical two-dimensional, infinitesimal area in
phase-space, then this area will grow exponentially with a rate determined by
the sum of the two largest Lyapunov exponents. In other words, the exponential
growth of a typical infinitesimal $n$-dimensional subvolume in phase space is
determined by the sum of the $n$ largest Lyapunov exponents. Further, for a
Hamiltonian system, the Lyapunov exponents come in plus-minus conjugate pairs,
so that the sum of a conjugate pair of exponents is always zero. Consequently,
the growth of an infinitesimal $dN$-dimensional subvolume is determined by the
sum of the $dN$  non-negative Lyapunov  exponents, and the volume of an
infinitesimal $2dN$-dimensional volume remains constant, in accord with
Liouville's theorem.

For the typical $dN$ dimensional subvolume, we consider a volume formed by the
projection of $2dN$ infinitesimal displacement vectors on velocity  space,
$(\vdv_1,\vdv_2,\ldots,\vdv_N)$.  Given  initial values for each of these
vectors, as well as for all of the positions, $\vec{r}_i$, velocities,
$\vec{v}_i$, and position deviation vectors, $\vdr_i$, we can follow their
evolution in time and, in principle at least, determine the time dependence of
an infinitesmal volume element in velocity space, which we denote as
$\delta{\cal V}(t)$. Then
\begin{eqnarray}
h_{KS} =\sum_{\lambda_{i}\geq 0}\lambda_{i} & = &
\lim_{t\to\infty}\frac{1}{t}\ln\frac{\delta{\cal V}(t)}{\delta{\cal
V}(0)} \nonumber \\
& = & \lim_{t\to\infty}\frac{1}{t}\int_0^{t}d\tau \frac{d\ln
\delta{\cal V}(\tau)}{d\tau} \nonumber \\
& = & \left\langle  \frac{d\ln
\delta{\cal V}(t)}{dt} \right\rangle.
\eqlabel{ks1}
\end{eqnarray}
The last line of \eq{ks1} is based on the assumption (still unproven) that a
gas of hard spheres is ergodic, so that time averages can be replaced by
equilibrium averages taken with respect to a microcanonical ensemble. Here this
average is denoted by angular brackets. Since  we are considering a volume
element in velocity space, we can use the fact that the velocity displacement
vectors do not change during the free flight motion of the particles between
the collisions, but do change at a collision. Under these circumstances, one
can use elementary kinetic theory considerations to show that the final term on
the right hand side of \eq{ks1} is
\begin{equation}
 \left\langle  \frac{d\ln
\delta{\cal V}(t)}{dt} \right\rangle =  \left\langle \sum_{i < j}{\cal
T}_{i,j}\ln \delta{\cal V}\right\rangle = \frac{N(N-1)}{2}\left\langle {\cal
T}_{12}\ln\delta{\cal V}\right\rangle.
\eqlabel{ks2}
\end{equation}
Here ${\cal T}_{12}$ is a binary collision operator, discussed in some detail
in \citerefs{hoegy,edjsp}, given by
\begin{equation}
{\cal T}_{12} =
\sigma^{d-1}\int_{\vec{v}_{12}\cdot\hat{\sigma}<0}
d\hat{\sigma} |\vec{v}_{12}\cdot\hat{\sigma}| \delta(\vec{r}_{12}-
\sigma\hat{\sigma})[{\cal
P}_{\hat{\sigma}}(1,2) -1].
\eqlabel{ks3}
\end{equation}
In \eq{ks3}, $d$ is the number of spatial dimensions of the system,  $\sigma$
again is the diameter of the spheres, $\vec{r}_{12}=
\vvr_1-\vvr_2;\vec{v}_{12} = \vec{v}_1 -\vec{v}_2$, and the operator ${\cal
P}_{\hat{\sigma}}(1,2)$ is a substitution operator which replaces the
precollision values, $\vvr_1,\vv_1,\vvr_2,\vv_2,\vdr_1,\vdv_1,\vdr_2,\vdv_2$,
by their post collision values, denoted with primes, given by
Eqs.~(\ref{eq:ll2},\ref{eq:ll3}). The  unit vector $\hat{\sigma}$ is an impact
parameter, running in the direction  of the line connecting the  centers at
collision and is integrated  over a hemisphere corresponding to all allowed
directions.

At this point it is useful to express the precollision quantities $\vdr_i $ as
\begin{equation}
\vdr_i =\vdr_i(0) + \tau_i
\vdv_i,
\eqlabel{ks4}
\end{equation}
where $\vdr_i(0)$ is the position displacement of particle $i$ just after its
previous collision, and $\tau_i$ is the time between the previous collision of
particle $i$ with some other particle, and the next collision involving
particle $i$. To further simplify the expression for the KS-entropy, we now
neglect, as in section \ref{sec:lyap}, the initial  displacement vectors,
$\vdr_i(0)$, when we calculate the change of the infinitesimal volume  in
velocity space at the $(1,2)$ collision in \eq{ks2}. This turns out to be a
serious approximation. It leads to the correct value for the leading density
term in $h_{KS}$, at low density, but the first order correction to this term
is obtained incorrectly in this approximation. This can be repaired, but at the
cost of a much longer and intricate calculation which we will present
elsewhere.

If we insert $\vdr_1 =  \tau_1\vdv_1$ and $\vdr_2 =  \tau_2\vdv_2$ into the
expression, \eq{ll2}, for the post collision  velocity deviations for particles
$1$ and $2$, we find that
\begin{equation}
\left[{\cal P}_{\hat{\sigma}}(1,2)-1\right]\ln\delta{\cal V}
=\ln\frac{\delta{\cal V}'}{\delta{\cal V}} = \ln |\det {\bf M}_{12}|,
\eqlabel{ks5}
\end{equation}
where
\begin{equation}
{\bf M}_{12} = {\bf 1}
-2\hat{\sigma}\hat{\sigma}-\frac{2T_{12}}{
\sigma}\left[(\vec{v}_{12}\cdot\hat{\sigma}){\bf
1} -\vec{v}_{12}\hat{\sigma} +\hat{\sigma}\vec{v}_{12}
-\frac{\vec{v}_{12}^{2}}{(\vec{v}_{12}
\cdot\hat{\sigma})}\hat{\sigma}\hat{\sigma
}\right].
\eqlabel{ks6}
\end{equation}
In \eq{ks6}, $T_{12} = ( \tau_1 +  \tau_2)/2$ and ${\bf 1}$ is the unit matrix.
The determinants are easily evaluated. For $d=2$, one finds
\begin{equation}
| \det {\bf M}_{12}| = 1 +\frac{2T_{12}|\vec{v}_{12}|}{
\sigma\cos\phi},
\eqlabel{ks7}
\end{equation}
where $\phi$ is the angle of incidence in the $1,2$ collision and ranges over
the values $-\pi/2 \leq \phi \leq \pi/2$. A similar calculation for $d=3$ shows
that
\begin{equation}
|\det{\bf M}_{12}| = 1 +
\frac{2T_{12}|\vec{v}_{12}|}{
\sigma\cos\phi}(\cos^{2}\phi +1)
+\left(\frac{2T_{12}|\vec{v}_{12}|}{
\sigma}\right)^{2}.
\eqlabel{ks8}
\end{equation}
To obtain the leading term in the density, for low density gases, we keep the
highest power of the time in each of the expressions for the determinant.
Further, at low densities we can compute the ensemble averages appearing in
\eq{ks1} by ignoring possible pre-collision correlations between particles $1$
and $2$, and  using equilibrium values for the single-particle distribution
functions appearing in the ensemble averages. In this way we find for $d=2$
\begin{eqnarray}
h_{KS}/N &=& \frac{a}{2n}\int d\vec{v}_1\int d\vec{v}_2
\int d\tau_1\int d\tau_2
\int_{\vec{v}_{12}\cdot\hat{\sigma}<0}
d\hat{\sigma}|\vec{v}_{12}\cdot\hat{\sigma}
|\times \nonumber \\ &&\times
F_1(\vec{v}_1,\tau_1)
F_1(\vec{v}_2,\tau_2
)\ln T_{12} +\cdots,
\eqlabel{ks9}
\end{eqnarray}
where the normalized equilibrium single particle distribution functions,
$F_1(\vec{v}_i,\tau_i)$ are given, in $d$ dimensions, by
\begin{equation}
F_1(\vec{v}_i,\tau_i)
= n(\frac{\beta
m}{2\pi})^{d/2}\nu(\vec{v}_i)e^{-\beta m
\vec{v}_{i}^{2}/2}e^{-\nu(\vec{v}_i)\tau_i}.
\eqlabel{ks10}
\end{equation}
Here $n$ is the number density of the gas, $\beta = (k_BT)^{-1}$, where $T$ is
the gas temperature, and $k_B$ is Boltzmann's constant, $\nu(\vec{v}_{i})$ is
the equilibrium collision frequency for a particle with velocity $\vec{v}_i$.
For  two dimensions, the evaluation of the  integrals leads directly to
\begin{equation}
h_{KS}/N = \frac{\nu}{2}[-\ln(n
\sigma^2) +\cdots]
\eqlabel{ks11}
\end{equation}
where $\nu =[(2\pi^{1/2}n \sigma)/(\beta m)^{1/2}]$ is the average collision
frequency at equilibrium for a two-dimensional gas of hard disks. The terms
left out are of higher order in the density.

For three-dimensional gases, a parallel calculation leads to
\begin{equation}
h_{KS}/N = \nu[-\ln(\pi n
\sigma^{3}) +\cdots],
\eqlabel{ks12}
\end{equation}
where for a gas of hard spheres ($d=3$) the average collision frequency $\nu
=[(4\pi^{1/2} n \sigma^{2})/(\beta m)^{1/2}]$.  These results are in excellent
agreement with the numerical simulations of Dellago and Posch\cite{Dellago3}.
The higher order terms take, as mentioned earlier, considerably more work, and
are discussed  elsewhere\cite{jrdtbp}.

\section{Conclusions and Outlook}
\label{sec:conclusion}

In the previous sections we have reviewed some of the ideas that motivate the
interest in the chaotic foundations of non-equilibrium processes in fluids. We
have provided an elementary discussion of transitive, hyperbolic dynamical
systems such as the baker and the Arnold cat maps to illustrate some of the
central notions and dynamical quantities.  We then turned to the applications
of kinetic theory to compute the largest Lyapunov exponent and the KS entropy
for a dilute gas of hard disks or hard spheres. The explicit results obtained
by these methods are in good agreement with the results of computer
simulations, and, apart from the corrections due to the velocity dependence of
the collision  frequency referred to at the end of section \ref{sec:lyap},
represent the present state of the art in the analytical calculation of chaotic
quantities for dilute gases with short range, repulsive forces. There are,
however, still many open problems which need solving. Here we mention a few of
them: \begin{enumerate}

\item We have a theory for the leading density behavior of the largest Lyapunov
exponent for a dilute gas of disks or spheres. We also have some understanding
of the number dependence of this quantity when we are not quite in the
thermodynamic limit. However we do not know much about the higher density
corrections to this exponent, nor do we know anything about the rest of the
Lyapunov spectrum, other than the KS entropy per particle (in the
thermodynamic limit). The determination of the complete spectrum would be quite
an accomplishment.

\item Recent results of Rom-Kedar and Turaev\cite{romked} imply that systems
with short range repulsive forces, other than hard disks or spheres, may not be
totally hyperbolic. Instead their phase spaces may have elliptic islands where
the motion is not chaotic. It would be interesting to know  first of all if
there are any experimental or theoretical consequences of the existence of
these elliptic regions for non equilibrium processes in real fluid systems and
also whether such elliptic islands will persist for arbitrarily large
energies.

\item One important application of the methods described here is to the
determination of the chaotic properties of thermostatted, driven systems for
which a  non-equilibrium steady state is reached and maintained. The general
properties of such systems are described in the books of Hoover\cite{hoover}
and of Evans and Morriss\cite{evmo}, and  a clear mathematical description  has
recently been given by  Ruelle\cite{ruelle}. Of special interest is the
perturbation of the Lyapunov spectrum produced by the thermostatted driving
field. For dilute, random Lorentz gases it has been possible to use kinetic
theory methods to determine the spectrum when the field is
small\cite{Beijerenetal}.  It would be worthwhile to extend these results to
larger fields and to gases where all of the particles are moving.

\item The escape-rate formalism described here has two drawbacks: (a) It is not
at all easy to describe the fractal repeller that forms in the phase-space of a
system with many degrees of freedom. (b) Even in those cases where the sum of
the positive Lyapunov exponents on the repeller  can  be calculated
analytically, the KS entropy is not yet directly accessible to analytic
methods. Instead one has to use the transport coefficients and the sum of the
positive Lyapunov exponents to infer the KS entropy of trajectories on the
repeller. It would be very helpful to have a better understanding of the
properties of high-dimensional repellers and to have an independent analytical
means to compute the KS entropy of trajectories on the repeller.

\item One of the  main goals of current research in this area is to obtain, if
possible, some deeper understanding of the dynamical basis of the laws of
irreversible thermodynamics. For two dimensional diffusive models based upon
the baker map, it has been possible to show that the laws of irreversible
thermodynamics result from a careful analysis of the fractal structures that
appear in the relevant phase spaces of these models when the systems are in
non-equilibrium steady states\cite{gasen,tvb,gildo}. The main physical idea is
that entropy is produced by the irreversible loss of information when changes
are taking place in a system on  very fine scales, beyond experimental
resolution. However, as has been emphasized by other authors\cite{rondoco},
these models may be too simple and/or the thermostats considered may be too
special to allow for any general conclusions to be drawn. This area of research
is active and many issues remain to be understood.

\item Our description of fluid systems as composed  of classical particles
interacting  through repulsive, short range forces is certainly incomplete.
Typical fluid systems are better modeled by short range forces with both
attractive and repulsive regions. We do not yet know what effects a more
careful analysis of interparticle forces will have on our picture of the
chaotic behavior of fluids. An even more serious problem is connected with our
use of classical mechanics to describe systems which are quantum mechanical in
nature. We have almost no understanding of how to correctly obtain a quantum
version of the classical chaotic picture of fluids, or even know for sure if
such a thing is possible.

\end{enumerate}

\section*{Acknowledgements}
We thank A.\ Latz, P.\ Gaspard, M.\ H.\ Ernst, and E.\
G.\ D.\ Cohen for many helpful conversations.  JRD acknowledges support  by the
National Science Foundation (USA) under grant NSF PHY-96-00428. HvB and RvZ are
supported by FOM, SMC and by the NWO Priority Program Non-Linear Systems, which
are financially supported by the "Nederlandse Organisatie voor
Wetenschappelijk Onderzoek (NWO)".

\newcommand{\emptybibitem}[1]{\bibitem{#1}\typeout{reference '#1' still
needs to be completed.}#1.}

\end{document}